\documentclass[11pt]{article}

\usepackage[final]{acl}

\usepackage{times}
\usepackage{latexsym}
\usepackage[T1]{fontenc}
\usepackage[utf8]{inputenc}
\usepackage{microtype}
\usepackage{inconsolata}

\usepackage{color,xcolor}
\usepackage{epsfig}
\usepackage{graphicx}
\usepackage{lipsum}
\usepackage{array}
\usepackage{booktabs}
\usepackage{colortbl}
\usepackage{multirow}
\usepackage{float}
\usepackage{caption}
\usepackage{adjustbox}
\usepackage{subcaption}
\usepackage{footnote}
\makesavenoteenv{tabular}
\makesavenoteenv{table}
\usepackage{makecell}
\usepackage{tablefootnote}
\usepackage{threeparttable}

\usepackage{amsmath,amsfonts,amssymb,bm}
\usepackage[super]{nth}

\usepackage{changepage}
\usepackage{extramarks}
\usepackage{lastpage}
\usepackage{setspace}
\usepackage{soul}
\usepackage{xspace}
\usepackage{wrapfig}

\usepackage{url}
\usepackage{hyperref}
\hypersetup{colorlinks=True, urlcolor=black}

\usepackage[ruled,vlined]{algorithm2e}
\usepackage{enumitem}
\usepackage{verbatim}
\usepackage{pifont}
\usepackage[acronyms]{glossaries}
\glsdisablehyper
\def\ourmodel{CLSP}
\def\ourdataset{FCaps}

\title{Towards Fine-Grained and Multi-Granular Contrastive Language-Speech Pre-training}

\author{
\textbf{Yifan Yang\textsuperscript{1}\thanks{Equal Contribution. This work was done during an internship at Tencent Hunyuan.}},
\textbf{Bing Han\textsuperscript{1}$\footnotemark[1]$},
\textbf{Hui Wang\textsuperscript{3}},
\textbf{Wei Wang\textsuperscript{1}},
\textbf{Ziyang Ma\textsuperscript{1}},
\textbf{Long Zhou\textsuperscript{2}\thanks{Project Leader\quad$^{\ddagger}$Corresponding Author}} \\
\textbf{Zengrui Jin\textsuperscript{4}},
\textbf{Guanrou Yang\textsuperscript{1}},
\textbf{Tianrui Wang\textsuperscript{5}},
\textbf{Xu Tan\textsuperscript{2}},
\textbf{Xie Chen\textsuperscript{1,6$\ddagger$}}
\\
\textsuperscript{1}Shanghai Jiao Tong University
\textsuperscript{2}Tencent Hunyuan
\textsuperscript{3}Nankai University \\
\textsuperscript{4}Tsinghua University
\textsuperscript{5}Tianjin University
\textsuperscript{6}Shanghai Innovation Institute
\\
\texttt{\{yifanyeung,chenxie95\}@sjtu.edu.cn}}

\begin{document}

\maketitle

\begin{abstract}
Modeling fine-grained speaking styles remains challenging for language-speech representation pre-training, as existing speech-text models are typically trained with coarse captions or task-specific supervision, and scalable fine-grained style annotations are unavailable. We present \ourdataset{}, a large-scale dataset with fine-grained free-text style descriptions, encompassing 47k hours of speech and 19M fine-grained captions annotated via a novel end-to-end pipeline that directly grounds detailed captions in audio, thereby avoiding the error propagation caused by LLM-based rewriting in existing cascaded pipelines. Evaluations using LLM-as-a-judge demonstrate that our annotations surpass existing cascaded annotations in terms of correctness, coverage, and naturalness. Building on \ourdataset{}, we propose \ourmodel{}, a contrastive language-speech pre-trained model that integrates global and fine-grained supervision, enabling unified representations across multiple granularities. Extensive experiments demonstrate that \ourmodel{} learns fine-grained and multi-granular speech-text representations that perform reliably across global and fine-grained speech-text retrieval, zero-shot paralinguistic classification, and speech style similarity scoring, with strong alignment to human judgments. Code and dataset are publicly available at \url{https://github.com/yfyeung/CLSP}.
\end{abstract}
\begin{figure}[t]
\centering
\includegraphics[width=\linewidth]{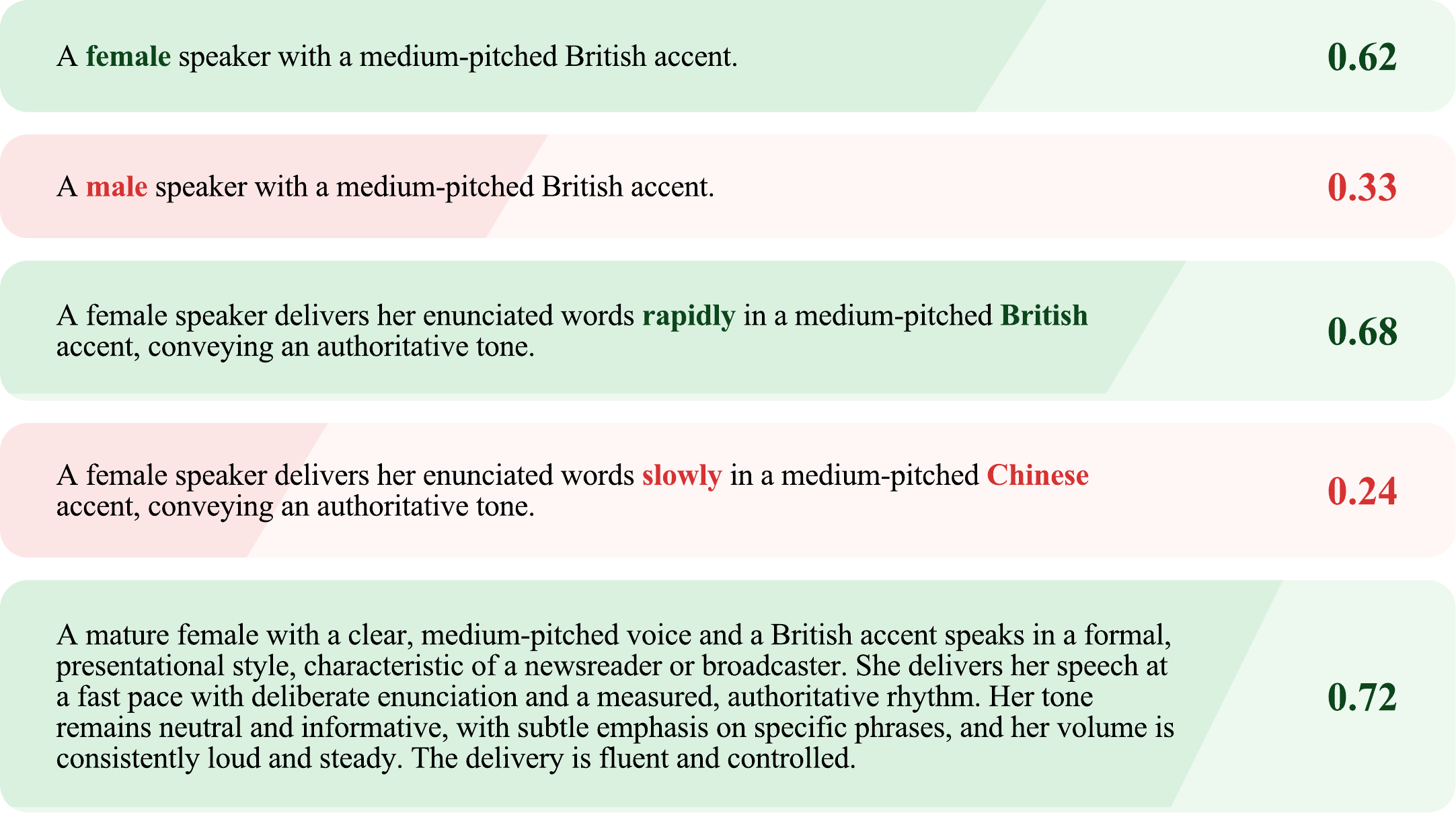}
\caption{Multi-granular speech style caption similarity scoring for the same speech input by \ourmodel{}.
Positive captions (green) receive higher scores, while hard negatives (red), despite mainly textual overlap, receive markedly lower scores due to attribute mismatches.}
\label{fig:demo}
\end{figure}

\definecolor{Green}{HTML}{2E7D32}
\definecolor{Red}{HTML}{C62828}

\begin{table*}[th]
\small
\centering
\caption{Comparison of \textbf{open-source} English speech style-captioned datasets. Following~\cite{capspeech}, I1-I5 denote intrinsic speaker traits: age (I1), gender (I2), timbre/texture (I3), mean pitch (I4), and accent (I5). S1-S4 represent situational traits: speaking rate (S1), emotion (S2), expressivity (S3), and volume (S4).}
\label{tab:related_dataset}
\renewcommand\tabcolsep{4pt}
\begin{tabular}{lcccccccccccc}
\toprule[1pt]
\textbf{Dataset}
&I1&I2&I3&I4&I5&S1&S2&S3&S4&Free-Text &End-to-End&\# Hours\\
\midrule
Expresso~\cite{expresso} &\textcolor{Green}{\ding{55}} &\textcolor{Green}{\ding{55}} &\textcolor{Green}{\ding{55}} &\textcolor{Green}{\ding{55}} &\textcolor{Green}{\ding{55}}&\textcolor{Green}{\ding{55}}&\textcolor{Red}{\ding{51}}&\textcolor{Red}{\ding{51}} &\textcolor{Green}{\ding{55}}&\textcolor{Green}{\ding{55}} &\textcolor{Green}{\ding{55}}&47\\

EARS~\cite{ears}&\textcolor{Red}{\ding{51}}&\textcolor{Red}{\ding{51}}&\textcolor{Green}{\ding{55}}&\textcolor{Red}{\ding{51}}&\textcolor{Red}{\ding{51}}&\textcolor{Red}{\ding{51}}&\textcolor{Red}{\ding{51}}&\textcolor{Red}{\ding{51}}&\textcolor{Red}{\ding{51}}&\textcolor{Green}{\ding{55}}&\textcolor{Green}{\ding{55}}&60 \\

PromptSpeech~\cite{prompttts} &\textcolor{Green}{\ding{55}}&\textcolor{Red}{\ding{51}} &\textcolor{Green}{\ding{55}}&\textcolor{Red}{\ding{51}} &\textcolor{Green}{\ding{55}}&\textcolor{Red}{\ding{51}}&\textcolor{Red}{\ding{51}}&\textcolor{Green}{\ding{55}}&\textcolor{Red}{\ding{51}}&\textcolor{Green}{\ding{55}}&\textcolor{Green}{\ding{55}}&0.3k\\

TextrolSpeech~\cite{textrolspeech} &\textcolor{Green}{\ding{55}}&\textcolor{Red}{\ding{51}}&\textcolor{Green}{\ding{55}}&\textcolor{Red}{\ding{51}}&\textcolor{Green}{\ding{55}}&\textcolor{Red}{\ding{51}} &\textcolor{Red}{\ding{51}}&\textcolor{Green}{\ding{55}}&\textcolor{Red}{\ding{51}}&\textcolor{Green}{\ding{55}}&\textcolor{Green}{\ding{55}}&0.3k\\

VccmDataset~\cite{controlspeech} &\textcolor{Green}{\ding{55}}&\textcolor{Red}{\ding{51}}&\textcolor{Green}{\ding{55}}&\textcolor{Red}{\ding{51}}&\textcolor{Green}{\ding{55}}&\textcolor{Red}{\ding{51}} &\textcolor{Red}{\ding{51}}&\textcolor{Green}{\ding{55}}&\textcolor{Red}{\ding{51}}&\textcolor{Green}{\ding{55}}&\textcolor{Green}{\ding{55}}&0.3k\\

LibriTTS-P~\cite{librittsp} &\textcolor{Red}{\ding{51}} &\textcolor{Red}{\ding{51}}&\textcolor{Red}{\ding{51}}&\textcolor{Red}{\ding{51}}&\textcolor{Green}{\ding{55}}&\textcolor{Red}{\ding{51}}&\textcolor{Green}{\ding{55}}&\textcolor{Red}{\ding{51}}&\textcolor{Red}{\ding{51}}&\textcolor{Green}{\ding{55}}&\textcolor{Green}{\ding{55}}&0.6k\\

DreamVoiceDB~\cite{dreamvoice} &\textcolor{Red}{\ding{51}} &\textcolor{Red}{\ding{51}} &\textcolor{Red}{\ding{51}} &\textcolor{Green}{\ding{55}} &\textcolor{Green}{\ding{55}} &\textcolor{Green}{\ding{55}} &\textcolor{Green}{\ding{55}}&\textcolor{Red}{\ding{51}} &\textcolor{Green}{\ding{55}}&\textcolor{Red}{\ding{51}}&\textcolor{Green}{\ding{55}}&0.3k\\

SpeechCraft~\cite{speechcraft} &\textcolor{Red}{\ding{51}}&\textcolor{Red}{\ding{51}}&\textcolor{Green}{\ding{55}}&\textcolor{Red}{\ding{51}} &\textcolor{Green}{\ding{55}}&\textcolor{Red}{\ding{51}}&\textcolor{Red}{\ding{51}}&\textcolor{Green}{\ding{55}}&\textcolor{Red}{\ding{51}}&\textcolor{Red}{\ding{51}}&\textcolor{Green}{\ding{55}}&2.4k\\

ParaSpeechCaps~\cite{paraspeechcaps} &\textcolor{Green}{\ding{55}}&\textcolor{Red}{\ding{51}}&\textcolor{Red}{\ding{51}}&\textcolor{Red}{\ding{51}}&\textcolor{Red}{\ding{51}}&\textcolor{Red}{\ding{51}}&\textcolor{Red}{\ding{51}}&\textcolor{Red}{\ding{51}}&\textcolor{Red}{\ding{51}}&\textcolor{Red}{\ding{51}}&\textcolor{Green}{\ding{55}}&2.9k\\

ParlerTTS~\cite{parlertts_dataset} &\textcolor{Green}{\ding{55}}&\textcolor{Red}{\ding{51}}&\textcolor{Green}{\ding{55}}&\textcolor{Red}{\ding{51}}&\textcolor{Red}{\ding{51}}&\textcolor{Red}{\ding{51}}&\textcolor{Green}{\ding{55}}&\textcolor{Red}{\ding{51}}&\textcolor{Green}{\ding{55}}&\textcolor{Red}{\ding{51}}&\textcolor{Green}{\ding{55}}&44.5k\\

CapSpeech~\cite{capspeech} &\textcolor{Red}{\ding{51}}&\textcolor{Red}{\ding{51}}&\textcolor{Red}{\ding{51}}&\textcolor{Red}{\ding{51}}&\textcolor{Red}{\ding{51}}&\textcolor{Red}{\ding{51}}&\textcolor{Red}{\ding{51}}&\textcolor{Red}{\ding{51}}&\textcolor{Red}{\ding{51}}&\textcolor{Red}{\ding{51}}&\textcolor{Green}{\ding{55}}&33.6k\\

\midrule
\textbf{\ourdataset} &\textcolor{Red}{\ding{51}}&\textcolor{Red}{\ding{51}}&\textcolor{Red}{\ding{51}}&\textcolor{Red}{\ding{51}}&\textcolor{Red}{\ding{51}}&\textcolor{Red}{\ding{51}}&\textcolor{Red}{\ding{51}}&\textcolor{Red}{\ding{51}}&\textcolor{Red}{\ding{51}}&\textcolor{Red}{\ding{51}}&\textcolor{Red}{\ding{51}}&47.1k\\
\bottomrule[1pt]
\end{tabular}
\end{table*}

\section{Introduction}
Speaking style conveys rich paralinguistic information beyond lexical content, encompassing both speaker-intrinsic characteristics (e.g., gender, age, and accent) and temporally varying traits (e.g., intonation, emotion, and expressiveness). In this sense, speaking style is inherently multi-scale: it can be summarized at the global, utterance level without explicit temporal structure, or described through fine-grained stylistic variation within an utterance.

However, modeling speaking style remains challenging. Existing approaches~\cite{emotion2vec} typically rely on utterance-level, discrete labels, which limit expressive diversity and fail to capture the temporal structure of speech.
Evaluating speaking style is likewise challenging. Widely used subjective human judgments suffer from limited inter-rater consistency and are hard to scale~\cite{emovoice}.
Recent automated alternatives based on large audio language models~\cite{gpt4o, Gemini25} incur huge costs.
As a result, there remains substantial scope for fine-grained modeling and scalable evaluation of speaking style in speech--text representation learning.

A central bottleneck is the lack of scalable, reliable, and fine-grained style annotations.
Existing speech style-captioned datasets~\cite{speechcraft, paraspeechcaps, capspeech} predominantly use cascaded annotation pipelines, in which speech is first labeled with discrete attributes and then rewritten into free-text descriptions by large language models, often introducing error propagation and semantic misalignment.
More fundamentally, these intermediate discrete attributes impose a severe information bottleneck, compressing rich, continuous, and temporally varying paralinguistic information into a finite set of predefined categorical tags and leading to substantial information loss.

With these perspectives in mind, we introduce \ourdataset{}, a large-scale speech dataset paired with fine-grained free-text speaking-style descriptions, constructed via an end-to-end annotation pipeline that directly grounds style captions in audio and incorporates agentic verification to improve annotation quality.
We show that our annotations achieve substantially higher quality than cascaded annotations in terms of correctness, coverage, and naturalness, as assessed by LLM-as-a-judge.
Building on \ourdataset{}, we further propose \ourmodel{}, a contrastive language--speech pre-trained model that leverages fine-grained and multi-granular supervision to learn unified speech--text representations across different levels of granularity (illustrated in Figure~\ref{fig:demo}).
Extensive experiments demonstrate strong performance across multiple tasks, including global and fine-grained speech--text retrieval, zero-shot paralinguistic classification, and speech style similarity scoring aligned with human judgments.

Our contributions are fourfold:
\begin{itemize}[leftmargin=*, itemsep=0pt, topsep=2pt]
\item We present \ourdataset{}\footnote{\url{https://huggingface.co/datasets/yfyeung/FCaps}}, the largest dataset to date for fine-grained free-text speaking-style descriptions, with 47k hours of speech and 19M captions.

\item We propose an end-to-end pipeline for fine-grained style annotation, substantially improving annotation quality over existing approaches.

\item We present \ourmodel{}\footnote{\url{https://huggingface.co/yfyeung/CLSP}}, a speech--text dual-encoder trained with fine-grained and multi-granular contrastive supervision, enabling unified representation learning across multiple granularities.

\item We demonstrate that \ourmodel{} learns robust and generalizable speech--language representations that perform reliably across multiple tasks, with strong alignment to human judgments.
\end{itemize}

\begin{figure*}[th]
    \centering
    \includegraphics[width=\linewidth]{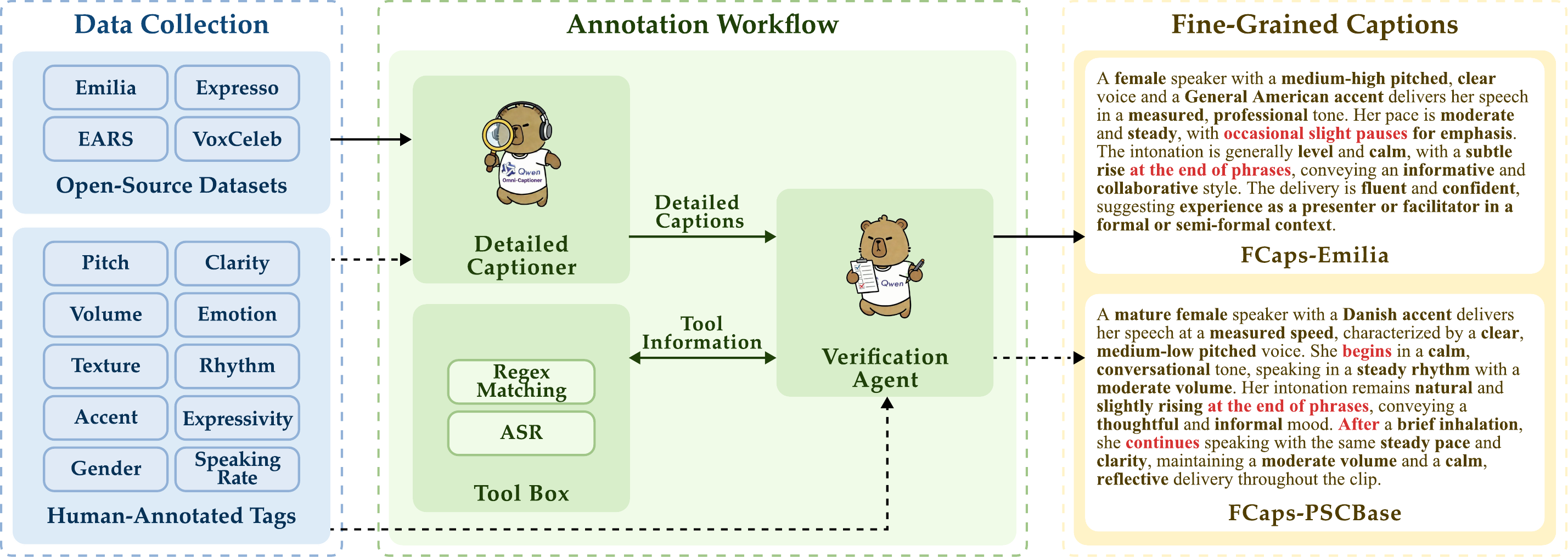}
    \caption{Overview of our end-to-end annotation pipeline for generating fine-grained captions, consisting of a detailed captioner and agentic verification with specialist tools. Solid lines indicate the construction process for \ourdataset{}-Emilia, and dashed lines indicate additional processes for \ourdataset{}-PSCBase. In the example fine-grained captions, speaker-related traits are highlighted in bold and narrative structure in red.}
    \label{fig:pipeline}
\end{figure*}

\section{Related Work}

\subsection{Speech Style-Captioned Datasets}
Early efforts such as EARS~\cite{ears} and Expresso~\cite{expresso} provide human-annotated discrete labels to describe speaking style, while NLSpeech~\cite{instructtts} employs annotators to describe speech emotion in natural language.
Subsequent works typically adopt a cascaded annotation pipeline, where speech is first labeled with discrete tags and then rewritten into natural-language captions.
LibriTTS-P~\cite{librittsp} extends LibriTTS-R~\cite{librittsr} by collecting perceptual and impression words and inserting them into predefined templates to form sentences.
DreamVoiceDB~\cite{dreamvoice} provides human-annotated voice timbre keywords, with corresponding descriptions generated by GPT-4~\cite{gpt4} based on their combinations.
ParaSpeechCaps~\cite{paraspeechcaps} extends EARS, VoxCeleb~\cite{voxceleb}, VoxCeleb2~\cite{voxceleb2}, and Expresso with additional human-annotated speaking style tags, which are then rewritten into natural language captions using Mistral-7B-Instruct-v0.2~\cite{mistral7b}.
CapSpeech~\cite{capspeech} similarly rewrites style tags from existing datasets using Mistral-7B-Instruct-v0.3~\cite{mistral7b}.
Such cascaded designs compress continuous paralinguistic cues into a finite set of discrete tags, introducing an information bottleneck that yields coarse captions and limits fine-grained modeling.
In contrast to existing speech style-captioned datasets summarized in Table~\ref{tab:related_dataset}, our approach adopts an end-to-end pipeline that avoids these limitations.

\subsection{Contrastive Language--Speech Pre-training}
Contrastive learning has emerged as a powerful paradigm for multimodal pre-training, demonstrating strong performance across image--text~\cite{clip} and audio--text~\cite{clap, laionclap}.
In the speech domain, existing speech--text contrastive models are predominantly trained with coarse-grained captions or task-specific supervision, offering limited modeling of the temporal and narrative structure of speech.
GLAP~\cite{glap} learns general representations across speech, audio, and music, but relies on paired transcriptions for speech, providing primarily lexical-level supervision.
RA-CLAP~\cite{raclap} adopts coarse-grained captions.
ParaCLAP~\cite{paraclap} focuses on emotion-centric supervision.
Overall, modeling fine-grained and multi-granular speaking styles in a unified manner remains underexplored.

\section{\ourdataset{} Dataset Construction}

\subsection{Caption Taxonomy: Global and Fine-Grained}
We define two types of textual supervision:
\begin{itemize}[leftmargin=*, itemsep=0pt, topsep=2pt]
\item \textbf{Global captions} provide an atemporal, utterance-level description of a speech clip that summarizes speaker-related attributes, encompassing intrinsic traits tied to a speaker’s identity and situational traits that may vary across utterances. Such captions collapse speech into a single holistic description without temporal or narrative structure.

\item \textbf{Fine-grained captions} extend beyond a holistic speaker profile by explicitly modeling within-utterance temporal dynamics. They provide temporal and narrative structure that tracks how vocal behaviors evolve over time, including style shifts, prosodic variations, emphasis patterns, and non-verbal vocalizations. Such captions can further encode the speaker’s delivery style, communicative role, and communicative intent.
\end{itemize}
Together, they provide multi-granular views of the same speech signal, thereby supporting fine-grained contrastive learning of a unified representation across multiple granularities.

\subsection{End-to-End Annotation Pipeline}
\label{sec:pipeline}
Figure~\ref{fig:pipeline} illustrates our proposed end-to-end annotation pipeline for generating fine-grained speech captions. 
Given a speech clip as input, the pipeline consists of two processes: detailed caption generation and agentic verification. 
A multimodal captioner generates multiple candidate captions for the input speech, optionally taking available human-annotated tags.
A verification agent then evaluates each candidate using a predefined checklist and a toolbox of specialist tools, and decides whether the candidate is retained or filtered.

\paragraph{Detailed Captioning}
Detailed captioning, also referred to as detailed perception, aims to generate fine-grained descriptions of an audio or video segment, emphasizing maximal perceptual detail within the clip duration.
In this work, we employ Qwen3-Omni-30B-A3B-Captioner\footnote{\url{https://huggingface.co/Qwen/Qwen3-Omni-30B-A3B-Captioner}}~\cite{qwen3omni} as the detailed captioner, which produces detailed and low-hallucination captions, capturing speaker emotions, layered intentions, cultural context, implicit cues, and additionally supports non-speech sound recognition and analysis.

Although the captioner is designed to operate without prompting, its default outputs often include spoken-content transcription, environmental sound descriptions, and audio quality assessments, which are irrelevant for speaker-centric contrastive learning.
We therefore apply user prompt conditioning to suppress such content. Leveraging the strong instruction-following capability inherited from the base model, the captioner can reliably adhere to these constraints, thereby steering caption generation towards relatively concise and speaker-focused descriptions.
A qualitative case study on how different prompt compositions influence captioner outputs is provided in Appendix~\ref{sec:user_prompt_detailed_caption}.

\paragraph{Multi-Positive Captioning}
To construct multiple positive textual views of the same speech clip, we perform caption generation multiple times using different random seeds.
This yields a set of captions that are all grounded in the same speech signal, while differing in lexical choice, descriptive focus, and narrative structure.
Previous work~\cite{speechcraft,capspeech} adopts text-based data augmentation that rewrites captions purely in the textual modality using LLMs with complex instructions. Such approaches suffer from instruction non-compliance and introduce distorted attributes or hallucinations in a non-trivial fraction of cases, as the rewriting process is not conditioned on the original speech signal.
By contrast, our approach conditions the generation process directly on the audio input, producing multiple acoustically consistent yet semantically non-identical positive views that are well suited for contrastive learning.
A case study is provided in Appendix~\ref{sec:multi_postive_caption}.

\paragraph{Agentic Verification}
To improve reliability of generated captions, we perform a verification process using the text-based reasoning model Qwen3-30B-A3B-Thinking-2507\footnote{\url{https://huggingface.co/Qwen/Qwen3-30B-A3B-Thinking-2507}}~\cite{qwen3}.

The verification agent evaluates each candidate caption according to a predefined checklist and discards it if \emph{any} item in the checklist is violated.
The checklist targets common failure modes in detailed captioning by examining whether a caption:
(1) includes descriptions of background sounds, environmental noise, or audio quality;
(2) contains explicit declarations about the absence of certain elements;
(3) contains spoken-content transcription without attached style descriptions; or
(4) fails to appropriately incorporate human-annotated tags when available.
In addition, for speech clips from EARS~\cite{ears}, Expresso~\cite{expresso}, and VoxCeleb~\cite{voxceleb}, the agent enforces a clip-level constraint requiring a single speaker with a single role. If a caption indicates multiple speakers or a single speaker assuming multiple roles, the corresponding speech clip and all associated captions are discarded.

To support these judgments, the agent leverages specialist tools, including (1) rule-based pattern matching implemented via Python regular expressions, (2) access to the speech transcription, and (3) access to human-annotated tags when available.
Based on the aggregated evidence, the agent makes a binary retain-or-filter decision.
This agentic verification process enables systematic quality control, effectively eliminating captions containing extraneous content or incomplete grounding.

\subsection{\ourdataset{}-Emilia}
\ourdataset{}-Emilia is constructed from the Emilia corpus~\cite{emilia}.
For each speech segment, the detailed captioner is run five times to generate candidate fine-grained captions.
Given the large scale of the audio sources, the verification agent retains a single verified fine-grained caption per utterance.
In total, \ourdataset{}-Emilia comprises 18,131,371 fine-grained captions, covering 46,787 hours of speech.
\ourdataset{}-Emilia does not include global captions.

\subsection{\ourdataset{}-PSCBase}
\ourdataset{}-PSCBase is built upon the PSC-Base corpus~\cite{paraspeechcaps}, incorporating audio clips from EARS, Expresso, and VoxCeleb, along with additional human-annotated tags and captions.
We adopt the captions provided by PSC-Base as global captions and apply rule-based normalization to mitigate common artifacts arising from LLM rewriting and fuzzy matching to correct spelling errors.
We incorporate speaking rate, accent, and situational tags to guide fine-grained caption generation.
For each utterance, the detailed captioner is run 20 times to obtain candidate fine-grained captions.
After agentic verification, between 5 and 14 verified captions per utterance are retained as multi-positive views.
Overall, \ourdataset{}-PSCBase comprises 140,602 global captions and 930,917 fine-grained captions, spanning 267 hours of speech.

\section{\ourmodel{} Model}

\subsection{Model Architecture}
\begin{figure}[t]
\centering
\includegraphics[width=\linewidth]{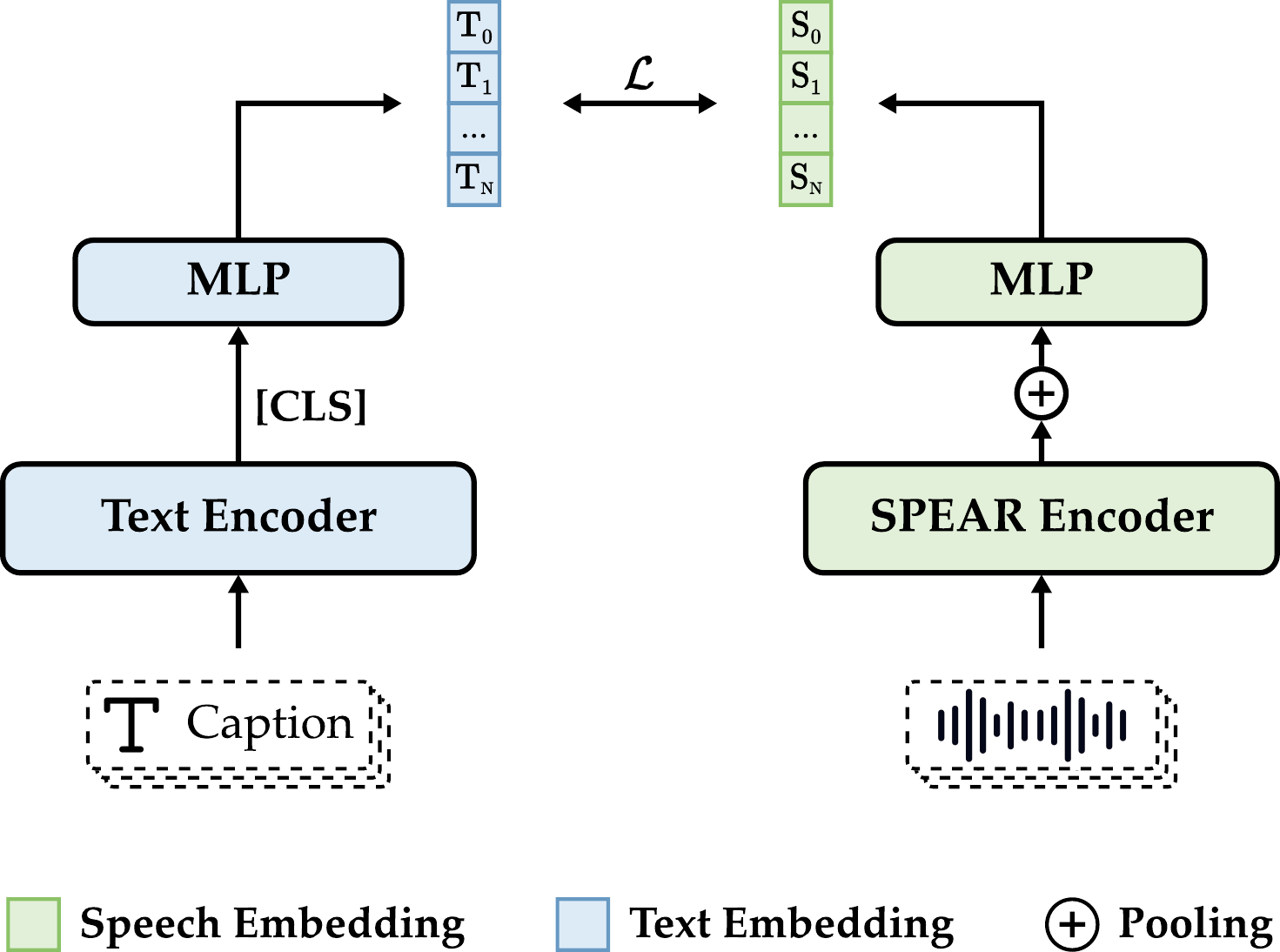}
\caption{Overview of \ourmodel{}.}
\label{fig:model}
\end{figure}

As shown in Figure~\ref{fig:model}, \ourmodel{} adopts the dual-encoder architecture of CLAP, where speech and text are processed by separate encoders, followed by MLP projection to map two modalities into a shared embedding space.

\paragraph{Speech and Audio Unified Encoder}
SPEAR-XLarge\footnote{\url{https://huggingface.co/marcoyang/spear-xlarge-speech-audio}} is used as the speech and audio unified encoder, with representations extracted from the final encoder layer.
SPEAR~\cite{spear} is a unified self-supervised representation model for both speech and general audio, and achieves state-of-the-art performance across a range of speech and audio benchmarks, making it well-suited for capturing fine-grained acoustic and paralinguistic cues in speaker-centric contrastive learning.

\paragraph{Text Encoder}
RoBERTa-base\footnote{\url{https://huggingface.co/FacebookAI/roberta-base}}~\cite{roberta} is used as the text encoder, with variable-length inputs up to 512 tokens to accommodate captions of varying granularity.
The sentence-level representation is taken from the final-layer \texttt{[CLS]} token.

\subsection{Fine-Grained and Multi-Granular Contrastive Language--Speech Pre-training}
We adopt a two-stage curriculum for speech--text representation learning with fine-grained and multi-granular contrastive supervision.
The training focus progressively shifts from pure fine-grained alignment to a balance between cross-granularity generalization and robust fine-grained discrimination.
In the first stage, speech and text are aligned using standard contrastive learning on large-scale data paired with fine-grained captions.
In the second stage, we introduce multi-positive contrastive learning with diverse global and fine-grained captions at multiple granularities.
We provide ablation studies on multi-stage training in~Appendix~\ref{sec:ablation_multi_stage_training}.

\paragraph{Stage One}
Given a speech clip $\mathbf{x}$ and its paired tokenized fine-grained caption $\mathbf{y}_{\mathrm{F}}$, the speech encoder produces frame-level representations that are aggregated via mean pooling over time, followed by MLP projection and $\ell_2$ normalization to obtain a speech embedding
$\mathbf{s} \in \mathbb{R}^d$.
For text, we take the final-layer \texttt{[CLS]} hidden state from the text encoder, followed by an MLP projection
and $\ell_2$ normalization to obtain a text embedding $\mathbf{t}_{\mathrm{F}} \in \mathbb{R}^d$. Here, $d$ is the embedding space dimensionality.
We use a symmetric InfoNCE~\cite{infonce} loss, with each paired speech and text forming a positive example,
and all other non-matching pairs within the same batch serving as negatives:
\begin{align}
\mathcal{L} = &-\frac{1}{2N}\sum_{i=1}^{N} \Bigg(
\log \frac{\exp(\mathbf{s}_i \cdot \mathbf{t}_{\mathrm{F}i}/\tau)}
{\sum_{j=1}^{N} \exp(\mathbf{s}_i \cdot \mathbf{t}_{\mathrm{F}j}/\tau)}
\notag\\
&+
\log \frac{\exp(\mathbf{t}_{\mathrm{F}i} \cdot \mathbf{s}_i/\tau)}
{\sum_{j=1}^{N} \exp(\mathbf{t}_{\mathrm{F}i} \cdot \mathbf{s}_j/\tau)}
\Bigg),
\end{align}
where $N$ is the batch size and $\tau$ is a learnable temperature. Since all embeddings are $\ell_2$-normalized, dot products are equivalent to cosine similarity.

\paragraph{Stage Two}
We adopt a symmetric multi-positive InfoNCE loss, implemented as cross-entropy with soft targets.
Given a batch of $N$ speech samples $\{\mathbf{x}_i\}_{i=1}^{N}$, each paired with two tokenized captions $\{\mathbf{y}_i, \hat{\mathbf{y}}_i\}$, we obtain a speech embedding $\mathbf{s}_i \in \mathbb{R}^d$ and two text embeddings $\mathbf{t}_i, \hat{\mathbf{t}}_i \in \mathbb{R}^d$ in the same manner as in Stage One.
We stack the speech embeddings as
$\mathbf{S}=[\mathbf{s}_1,\ldots,\mathbf{s}_N] \in \mathbb{R}^{N\times d}$
and the text embeddings as
$\mathbf{T}=[\mathbf{t}_1,\ldots,\mathbf{t}_N,\hat{\mathbf{t}}_1,\ldots,\hat{\mathbf{t}}_N] \in \mathbb{R}^{2N\times d}$, and compute the similarity logits $\mathbf{L}=\mathbf{S}\mathbf{T}^{\top} \in \mathbb{R}^{N\times 2N}$.
For audio-to-text direction, we define a soft target distribution $\mathbf{D} \in \mathbb{R}^{N \times 2N}$ that assigns probability mass $\lambda$ and $1-\lambda$ to two paired texts, and zero to all others:
\begin{equation}
D_{i,j} =
\begin{cases}
\lambda, & \text{if } j = i,\\
1-\lambda, & \text{if } j = i+N,\\
0, & \text{otherwise}.
\end{cases}
\label{eq:multi_postive_target}
\end{equation}
We set $\lambda=0.5$ based on the ablation study reported in Appendix~\ref{sec:ablation_target_weight}.
For text-to-audio direction, each text embedding has a single speech, yielding target distribution $\mathbf{D'} \in \mathbb{R}^{2N \times N}$:
\begin{equation}
D'_{j,i} =
\begin{cases}
1, & \text{if } j = i \text{ or } j = i+N,\\
0, & \text{otherwise}.
\end{cases}
\end{equation}
The loss is defined as the average of two directions:
\begin{equation}
\mathcal{L} =
\frac{1}{2}\Big(
\mathrm{CE}\big(\mathbf{L}/\tau, \mathbf{D}\big)
+
\mathrm{CE}\big(\mathbf{L}^{\top}/\tau, \mathbf{D}'\big)
\Big),
\end{equation}
where $\mathrm{CE}(\cdot,\cdot)$ denotes cross-entropy and $\tau$ is a learnable temperature parameter.

For each training step, the model samples one of two tasks according to a task scheduler:
\begin{itemize}[leftmargin=*, itemsep=0pt, topsep=2pt]
\item \textbf{Task~1}: each speech sample is paired with a global caption and a fine-grained caption, encouraging cross-granularity generalization.
\item \textbf{Task~2}: each speech sample is paired with two distinct fine-grained captions, improving fine-grained discrimination via semantic consistency.
\end{itemize}
We explore both static and dynamic task schedulers.
At training step $t$, Task~1 is sampled with probability $p_t$, while Task~2 is sampled with probability $1-p_t$. For the static scheduler, $p_t=p_0$ is fixed. For the dynamic scheduler, $p_t$ decreases linearly from $p_0$ to $p_{\min}$ over $T$ training steps and remains fixed thereafter:
\begin{equation}
p_t = \max\!\left( p_{\min},\; p_0 - \frac{t}{T}(p_0 - p_{\min}) \right).
\end{equation}
The best-performing strategy uses a dynamic scheduler with $p_0=0.95$, $p_{\min}=0.50$, and $T=10{,}000$.
We provide ablation studies of different task schedulers in~Appendix~\ref{sec:ablation_task_scheduler}.
\section{Experiments}
\subsection{Annotation Quality of \ourdataset{}}

\paragraph{Evaluation Setup}
We evaluate the annotation quality of \ourdataset{} by comparing captions generated by our end-to-end annotation pipeline with those produced by a representative cascaded annotation pipeline.
Specifically, we randomly sample 1,000 audio clips from \ourdataset-Emilia along with one caption per audio generated by our end-to-end pipeline.
For the same set of audio clips, we obtain the corresponding captions from PSC-Scaled~\cite{paraspeechcaps} as the cascaded baseline, which consists of automatically predicted discrete style labels, filtered by Gemini~1.5 Flash~\cite{Gemini15}, and rewritten into natural-language style captions by Mistral-7B-Instruct-v0.2~\cite{mistral7b}.

\paragraph{LLM-as-Judges}
We use gemini-3-pro-preview, a natively multimodal LLM~\cite{gemini}, as a judge to evaluate caption quality.
Each evaluation query consists of an audio clip and its two corresponding captions (cascaded \textit{vs.} end-to-end), and assigns absolute scores to each caption along three dimensions:
(1) audio-grounded \emph{correctness}, measuring factual consistency with the audible content;
(2) \emph{coverage}, assessing whether speaking-style attributes are adequately captured;
and (3) \emph{naturalness}, evaluating fluency, grammaticality, and human-likeness.
All evaluations follow an identical prompt and scoring rubric, detailed in Appendix~\ref{sec:llm_protocol}.
Scores are averaged over five runs with randomized ordering to reduce variance.

\begin{figure}[!t]
\centering
\includegraphics[width=\linewidth]{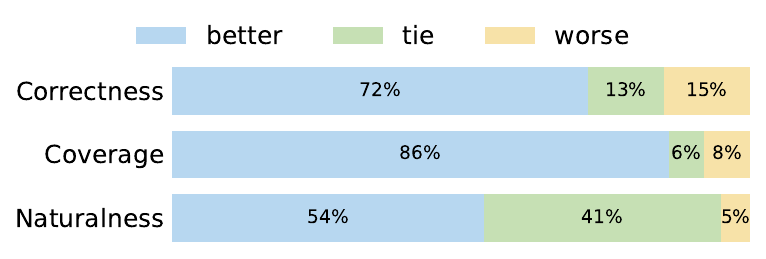}
\caption{
Pairwise comparison between end-to-end and cascaded captions across correctness, coverage, and naturalness dimensions, showing the proportions of better, tied, and worse cases under Gemini~3~Pro evaluation.
}
\label{fig:pairwise_analysis}
\end{figure}

\paragraph{Results}
Table~\ref{tab:gemini_score} reports Gemini~3~Pro scores for cascaded and end-to-end caption annotations.
Our end-to-end annotations consistently outperform cascaded annotations across all three dimensions by a large margin.
Figure~\ref{fig:pairwise_analysis} further shows that end-to-end captions outperform cascaded captions in the majority of cases across all evaluation dimensions, with especially large gains in coverage and correctness.
These gaps highlight the information bottleneck introduced by intermediate discrete attributes in cascaded pipelines, which irreversibly compress rich paralinguistic information.
By contrast, our end-to-end pipeline yields better alignment with the audible content, more comprehensive coverage of speaking-style attributes, and more fluent, human-like descriptions.

\begin{table}[!t]
\small
\centering
\caption{Comparison of end-to-end and cascaded caption annotations evaluated by Gemini~3~Pro.}
\label{tab:gemini_score}
\renewcommand\tabcolsep{1pt}
\begin{tabular}{lcccc}
\toprule[1pt]
\textbf{Pipeline}
& \textbf{Correctness}
& \textbf{Coverage}
& \textbf{Naturalness}
& \textbf{Avg.} \\
\midrule
Cascaded
& $3.30_{\pm 0.05}$
& $3.10_{\pm 0.02}$
& $4.15_{\pm 0.05}$
& $3.51_{\pm 0.04}$ \\
End-to-end
& $\mathbf{4.42}_{\pm 0.04}$
& $\mathbf{4.55}_{\pm 0.03}$
& $\mathbf{4.92}_{\pm 0.02}$
& $\mathbf{4.63}_{\pm 0.03}$ \\
\bottomrule[1pt]
\end{tabular}
\end{table}
\begin{table*}[th]
\small
\centering
\caption{Global speech--text retrieval and fine-grained speech--text retrieval results. Baselines are evaluated using public checkpoints. Best results are in \textbf{bold}.}
\label{tab:retrieval}
\begin{tabular}{lcccccccc}
\toprule[1pt]
\multirow{2}{*}{\textbf{System}}
& \multicolumn{4}{c}{\textbf{Speech-to-Text}}
& \multicolumn{4}{c}{\textbf{Text-to-Speech}} \\
\cmidrule(lr){2-5} \cmidrule(lr){6-9}
& R@1 & R@5 & R@10 & mAP@10
& R@1 & R@5 & R@10 & mAP@10 \\
\midrule

\multicolumn{9}{l}{\hspace{-2mm}\emph{Global Speech--Text Retrieval}} \\
LAION-AI CLAP~\cite{laionclap} & 0.4 & 2.5 & 5.4 & 1.5 & 0.8 & 3.3 & 5.0 & 1.9 \\
GLAP~\cite{glap} & 1.7 & 5.0 & 9.5 & 3.4 & 1.7 & 5.8 & 10.8 & 3.9 \\
ParaCLAP~\cite{paraclap} & 2.1 & 4.6 & 9.1 & 3.5 & 0.4 & 5.0 & 7.9 & 2.3 \\
\ourmodel & \textbf{45.6} & \textbf{75.9} & \textbf{84.2} & \textbf{58.7} & \textbf{40.3} & \textbf{74.3} & \textbf{82.6} & \textbf{54.5} \\
\midrule

\multicolumn{9}{l}{\hspace{-2mm}\emph{Fine-Grained Speech--Text Retrieval}} \\
LAION-AI CLAP~\cite{laionclap} & 0.4 & 3.3 & 5.4 & 1.6 & 1.2 & 2.5 & 2.9 & 1.7 \\
GLAP~\cite{glap} & 4.6 & 14.5 & 21.2 & 9.1 & 2.1 & 7.5 & 13.3 & 4.4 \\
ParaCLAP~\cite{paraclap} & 1.2 & 7.9 & 12.5 & 4.1 & 1.2 & 5.8 & 10.0 & 3.3 \\
\ourmodel & \textbf{68.1} & \textbf{90.9} & \textbf{95.9} & \textbf{77.9} & \textbf{67.2} & \textbf{90.9} & \textbf{96.3} & \textbf{77.2} \\

\bottomrule[1pt]
\end{tabular}
\end{table*}
\begin{table*}[th]
\small
\centering
\caption{
Zero-shot classification results, reported as WA (\%) / UA (\%).
$^{\dagger}$ indicates results taken from prior work; the others are evaluated using public checkpoints.
Best results are in \textbf{bold}.
}
\label{tab:zero-shot}
\begin{tabular}{l c@{ / }c c@{ / }c c@{ / }c c@{ / }c c@{ / }c}
\toprule[1pt]
\multirow{2}{*}{\textbf{System}}
& \multicolumn{2}{c}{\textbf{Emotion}} 
& \multicolumn{2}{c}{\textbf{Emotion}}
& \multicolumn{2}{c}{\textbf{Emotion}} 
& \multicolumn{2}{c}{\textbf{Gender}}
& \multicolumn{2}{c}{\textbf{Age}} \\
\cmidrule(lr){2-3} \cmidrule(lr){4-5} \cmidrule(lr){6-7} \cmidrule(lr){8-9} \cmidrule(lr){10-11}
& \multicolumn{2}{c}{IEMOCAP} 
& \multicolumn{2}{c}{RAVDESS} 
& \multicolumn{2}{c}{CREMA-D} 
& \multicolumn{2}{c}{RAVDESS}
& \multicolumn{2}{c}{CREMA-D} \\
\midrule
LAION-AI CLAP~\cite{laionclap} & 32.6 & 29.1 & 14.4 & 13.5 & 21.0 & 19.2 & 58.6 & 58.6 & 1.3 & 1.8 \\
GLAP~\cite{glap} & 32.5 & 27.0 & 7.8 & 13.0 & 13.6 & 17.9 & 72.6 & 72.6 & 27.1 & 32.6 \\
ParaCLAP~\cite{paraclap} & 46.1 & 46.5 & 28.1 & 30.3 & 29.8 & 29.6 & 99.2 & 99.2 & 31.2 & 32.0 \\
Auden-Voice CLAP~\cite{audenvoice}$^{\dagger}$ & -- & -- & 32.4 & -- & 30.2  & -- & 95.6 & -- & 38.5 & -- \\
\ourmodel & \textbf{57.2} & \textbf{56.1} & \textbf{46.8 }& \textbf{46.0} & \textbf{35.1} & \textbf{37.2} & \textbf{100.0} & \textbf{100.0} & \textbf{40.6} & \textbf{44.5}\\
\bottomrule[1pt]
\end{tabular}
\end{table*}
\begin{table*}[th]
\small
\centering
\caption{The correlation between the model-derived similarity scores and subjective MOS. Reported as Pearson / Spearman / Kendall's Tau correlation coefficients ($r, \rho, \tau$). All results are statistically significant with $p<0.01$.}
\label{tab:sub_score}
\renewcommand\tabcolsep{8pt}
\begin{tabular}{l c@{ / }c@{ / }c c@{ / }c@{ / }c c@{ / }c@{ / }c}
\toprule[1pt]
\textbf{System}
& \multicolumn{3}{c}{\textbf{Intrinsic Traits}} 
& \multicolumn{3}{c}{\textbf{Situational Traits}} 
& \multicolumn{3}{c}{\textbf{Fusion}} \\
\midrule
LAION-AI CLAP~\cite{laionclap} & 0.679 & 0.664 & 0.467 & 0.194 & 0.184 & 0.126 & 0.588 & 0.597 & 0.425 \\
GLAP~\cite{glap} & 0.372 & 0.340 & 0.234 & 0.169 & 0.138 & 0.102 & 0.350 & 0.344 & 0.241 \\
ParaCLAP~\cite{paraclap} & 0.663 & 0.634 & 0.445 & 0.323 & 0.330 & 0.232 & 0.626 & 0.599 & 0.417 \\
\ourmodel & \textbf{0.893} & \textbf{0.858} & \textbf{0.668} & \textbf{0.903} & \textbf{0.878} & \textbf{0.694} & \textbf{0.886} & \textbf{0.858} & \textbf{0.670} \\
\bottomrule[1pt]
\end{tabular}
\end{table*}

\subsection{\texorpdfstring{\ourmodel{}}{S3CLAP} as a Speech Task Evaluator}

\subsubsection{Implementation Details}
\ourmodel{} has 724M parameters in total, with 599M from SPEAR-XLarge and 125M from RoBERTa-base, trained on 8 NVIDIA A100 80GB GPUs, with a batch duration of 800 seconds per GPU.
The first stage runs for 1.2M steps, followed by 4k fine-tuning steps for the second stage, using ScaledAdam~\cite{zipformer} optimizer and Eden~\cite{zipformer} scheduler with peak learning rates of 0.045 and 0.001, respectively.

\subsubsection{Speech--Text Retrieval: Global and Fine-Grained}
\paragraph{Evaluation Setup}
We construct our test set using 241 audio clips from the ParaSpeechCaps test~\cite{paraspeechcaps} split, with durations ranging from 1 to 30 seconds, ensuring no overlap between training and test sets.
Global captions are derived from the PSC-Base holdout split, and fine-grained captions are generated by our end-to-end pipeline.

\paragraph{Evaluation Metrics}
We use standard metrics, including Recall at rank 1/5/10 (R@1, R@5, R@10) and mean Average Precision at 10 (mAP@10).
R@$k$ measures the proportion of queries whose correct match appears in the top-$k$, while mAP@10 measures ranking quality within the top 10.

\paragraph{Baselines}
As the proposed tasks are novel, we evaluate representative open-source audio--text retrieval models, including LAION-AI CLAP~\cite{laionclap}, GLAP~\cite{glap}, and ParaCLAP~\cite{paraclap}, which are pre-trained on coarse-grained or task-specific supervision without fine-grained captions supervision.
These models therefore serve to quantify the capability gap between existing approaches and our fine-grained modeling framework.
Detailed descriptions of the baseline systems are provided in Appendix~\ref{sec:baseline}.

\paragraph{Results}
Table~\ref{tab:retrieval} reports retrieval performance on both global and fine-grained tasks.
Across all settings, \ourmodel{} consistently outperforms all baselines by a large margin across all evaluation metrics for both speech-to-text and text-to-speech retrieval, while the baselines perform close to random guessing.
This substantial margin highlights the capability gap between existing models and \ourmodel{}, and underscores the effectiveness of training with global and fine-grained captions from \ourdataset{}.

\subsubsection{Zero-shot Paralinguistic Classification}
\paragraph{Evaluation Setup}
This task evaluates \ourmodel{}’s ability to recognize paralinguistic attributes under diverse attribute sets, without any task-specific training.
We focus on three representative paralinguistic dimensions: \textit{emotion}, \textit{gender}, and \textit{age}.
For each dimension, zero-shot classification is performed by computing the cosine similarity between a speech embedding and a set of natural-language text prompts describing candidates (e.g., ``A speaker in a happy tone.'', ``A male speaker.'', and ``A middle-aged speaker.''), and selecting the prompt with the highest similarity.

Our evaluations contain the following datasets:
\begin{itemize}[leftmargin=*, itemsep=0pt, topsep=2pt]
\item \textbf{IEMOCAP}~\cite{iemocap}: We use the 4-class emotion setup (happy/excited, angry, sad, neutral) with a total of 5,531 utterances.

\item \textbf{RAVDESS}~\cite{ravdess}: We use the speech part with 1,440 utterances, 8-class emotion categories (calm, happy, sad, angry, fearful, surprise, disgust, and neutral), and gender.

\item \textbf{CREMA-D}~\cite{cremad}: 7,442 utterances from 91 actors, with 6-class emotion categories (happy, sad, angry, fear, disgust, neutral) and speaker age, which we grouped into four bins (child, young adult, middle-aged, older).
\end{itemize}

\paragraph{Evaluation Metrics}
We report weighted accuracy (WA) and unweighted accuracy (UA), measuring overall and mean class accuracy, respectively.

\paragraph{Baselines}
We compare \ourmodel{} with strong baselines, including LAION-AI CLAP, GLAP, ParaCLAP, and Auden-Voice CLAP. 
For Auden-Voice CLAP, we use the ASR-pretrained variant without any supervised training on IEMOCAP, RAVDESS, or CREMA-D, ensuring strictly zero-shot evaluation.
Baseline details are provided in Appendix~\ref{sec:baseline}.

\paragraph{Results}
Table~\ref{tab:zero-shot} reports zero-shot paralinguistic classification results on emotion, gender, and age recognition.
Overall, \ourmodel{} consistently outperforms all baselines across datasets and paralinguistic dimensions.
These results indicate that \ourmodel{} learns generalizable speech representations that capture diverse paralinguistic semantics without task-specific supervision.

\subsubsection{Speech Style Similarity Scoring with Human Correlation}
\paragraph{Subjective Evaluation Setup}
To validate the reliability of \ourmodel{} as an automated metric for assessing speech-text alignment, we investigate its consistency with human perception for three distinct paralinguistic dimensions. We conduct a meta-evaluation using the holdout split of ParaSpeechCaps. The evaluation focuses on \emph{Intrinsic Traits}, \emph{Situational Traits}, and their \emph{Fusion}.

For human annotations, we recruited 20 experts with research backgrounds in speech processing to rate the matching degree between the audio and its corresponding text caption on a continuous scale.
We then calculate the correlation between these subjective scores and model-predicted similarity scores across several baselines, using three standard statistical coefficients: the Pearson Correlation Coefficient ($r$), Spearman’s Rank Correlation ($\rho$), and Kendall’s Tau ($\tau$). Details of the subjective evaluation are provided in Appendix~\ref{sec:subjective_eval_detail}.

\paragraph{Baselines}
We continue to use open-source representative audio-text models, including LAION-AI CLAP, GLAP, and ParaCLAP as baseline systems.

\paragraph{Results}
The experimental results, as summarized in Table~\ref{tab:sub_score}, demonstrate that \ourmodel{} significantly outperforms all baseline models in mirroring human perception across both intrinsic and situational traits, as well as their fusion.
Furthermore, the consistent performance across all evaluated metrics ($r, \rho, \text{ and } \tau$) confirms that \ourmodel{} not only tracks the absolute semantic matching quality linearly but also effectively preserves the relative ordinal ranking of samples in a manner that resonates with human expert judgment.
In conclusion, these results validate \ourmodel{} as a robust and high-fidelity automated proxy for human subjective evaluation in paralinguistic audio-text matching tasks. For a more intuitive assessment, visual comparisons highlighting the strong concordance between model-predicted similarity scores and human subjective ratings are provided in Appendix~\ref{sec:visual_subjective_evaluation}.

\paragraph{Discussion}
The strong alignment with human judgments suggests that \ourmodel{} can serve not only as a representation model but also as a human-aligned automatic metric~\cite{yang2026dswed, yang2026position} for speech generation tasks.
In particular, it naturally measures the capability of instruction-following TTS systems~\cite{instructtts}, where the goal is to generate speech that matches a natural language style prompt.
Compared to LLM-as-a-judge approaches~\cite{wang2025enablingauditoryllms, qualispeech, speechLLM_as_judges}, which rely on large audio language models and incur high computational cost, \ourmodel{} provides a scalable and practical alternative, making it well-suited for large-scale benchmarking and speech data filtering.
\section{Conclusion}
In this paper, we introduce \ourdataset{}, the largest speech dataset paired with fine-grained free-text speaking-style descriptions, together with \ourmodel{}, which learns fine-grained speech--text representations across multiple stylistic granularities.
Extensive experiments demonstrate that \ourmodel{} performs reliably across a range of tasks, including global and fine-grained speech--text retrieval, zero-shot paralinguistic classification, and speech style similarity scoring that aligns strongly with human judgments.
We hope this work encourages a shift in speaking style modeling from predefined, discrete attributes toward open-vocabulary, cross-granular, and speech-grounded natural language descriptions, facilitating more flexible and general-purpose speech--language representations.

\section*{Limitations}
While our approach demonstrates strong performance across a range of tasks, it has several limitations.
First, \ourmodel{} relies on a pre-trained speech and audio unified encoder that is trained only on English speech, and therefore our model is limited to English.
Second, publicly available paralinguistic speech data with diverse style attributes remains limited, particularly in its coverage of underrepresented accents, emotional expressions, and expressive speaking styles.
We leave extensions toward more diverse and multilingual speech representations as an important direction for future work.

\section*{Ethics Statement}
All data used in this work were collected and processed in accordance with relevant ethical guidelines and licensing terms.
Human annotations were performed by trained annotators, who were fairly compensated.
For \ourdataset{}, the speech samples are sourced from publicly available datasets. We release only annotated captions and associated metadata referencing the original audio under the CC BY-NC-SA 4.0 license and do not redistribute the original audio files.
\ourmodel{} is made publicly available under the Apache 2.0 license to support reproducibility and facilitate future research.
\section*{Acknowledgments}
This work was supported by the National Natural Science Foundation of China (No. U23B2018), Shanghai Municipal Science and Technology Major Project under Grant 2021SHZDZX0102, Yangtze River Delta Science and Technology Innovation Community Joint Research Project (2024CSJGG1100), and CIE-Tencent Doctoral Research Incentive Project.

\bibliography{custom}

\appendix
\section{Baseline Details}
\label{sec:baseline}

\begin{itemize}[leftmargin=*, itemsep=0pt, topsep=2pt]
    \item \textbf{LAION-AI CLAP}~\cite{laionclap}: An open-source\footnote{\url{https://github.com/LAION-AI/CLAP}} CLIP-style dual-encoder model with 158M parameters. 
    The audio encoder uses pre-trained HTSAT-tiny, with audio representations extracted from the penultimate layer before projection. 
    The text encoder uses RoBERTa-base, with text representation taken from the final-layer [CLS] token.
    LAION-AI CLAP is trained on large-scale audio-text paired data, including AudioCaps, Clotho, and the LAION-Audio-630K dataset.
    Training uses bidirectional contrastive learning with a symmetric cross-entropy loss.

    \item \textbf{GLAP}~\cite{glap}: An open-source\footnote{\url{https://github.com/xiaomi-research/dasheng-glap}} dual-encoder CLAP-style model with 855M parameters that aims to learn unified audio-text representations across speech, sound, and music domains. 
    GLAP employs a pre-trained general-purpose audio encoder and a multilingual text encoder based on Sonar for text representations. 
    GLAP is trained on large-scale audio-text paired data spanning speech (439.4M, 411k hours), sound (5.9M, 23.8k hours), and music (3k, 19.3 hours).
    Training uses a bidirectional contrastive objective based on a sigmoid loss.

    \item \textbf{ParaCLAP}~\cite{paraclap}: An open-source\footnote{\url{https://github.com/KeiKinn/ParaCLAP}} dual-encoder CLAP-style model with 276M parameters. 
    The speech encoder is initialized from a pruned version of Wav2Vec2-Large-Robust that has been further fine-tuned for dimensional speech emotion recognition on MSP-Podcast. 
    Audio representations are obtained by pooling the hidden states of the final Transformer layer. 
    The text encoder is bert-base-uncased, and the text representation is taken from the final-layer [CLS] token. 
    ParaCLAP is trained on the MSP-Podcast dataset, which contains nine emotion categories, utilizing bidirectional contrastive learning with a symmetric cross-entropy loss.

    \item \textbf{Auden-Voice CLAP}~\cite{audenvoice}: A dual-encoder CLAP-style model with 281M parameters, which remains closed-source at the time of submission. 
    The speech encoder is initialized from an ASR-pretrained Zipformer-L encoder trained on large-scale in-house Chinese ASR data using the RNNT loss. 
    Audio representations are obtained by mean-pooling the hidden states of the final Transformer layer. 
    The text encoder is RoBERTa-base, and text representations are taken from the final-layer [CLS] token.
    Different continual pre-training settings are applied starting from the ASR-pretrained encoder, including speaker identification trained on VoxCeleb2 (974k utterances and 2,026 hours), paralinguistic attribute prediction trained on CREMA-D, RAVDESS, IEMOCAP, and TESS (18.3k utterances and 20 hours), and a multi-task setting combining speaker identification and paralinguistic tasks. 
    Encoders obtained from each setting are used to initialize different CLAP variants, which are further trained on ParaSpeechCaps using both PSC-Base and PSC-Scaled splits (1.036M audio-text pairs, 2,700 hours) with bidirectional speech--text contrastive learning.
\end{itemize}

\definecolor{Green}{HTML}{2E7D32}
\definecolor{Red}{HTML}{C62828}

\begin{table*}[!t]
\small
\centering
\caption{Case study of different user prompt compositions and corresponding generated fine-grained captions.}
\label{tab:case_study_user_prompt}
\setlength{\tabcolsep}{4pt}
\begin{tabular}{p{0.26\linewidth} p{0.70\linewidth}}
\toprule[1pt]
\textbf{User Prompt Composition} & \textbf{Fine-grained Caption} \\
\midrule
Audio-only & The audio clip begins abruptly, mid-sentence, as a male speaker with a distinct Afro-Caribbean accent continues a thought: \textcolor{Red}{“…of of the idea that has been the same idea for a thousand years that they believe that—”}. His delivery is measured and analytical, marked by a slight hesitation and repetition at the start, indicating an unscripted, conversational tone. The speaker’s voice is steady, low-to-mid in pitch, and slightly gravelly, with clear articulation and no signs of emotional distress or heightened affect. \textcolor{Red}{The environment is quiet, featuring only a faint electronic hum and minimal room reverberation, pointing to a small, acoustically treated indoor setting such as a home studio or office.} \textcolor{Red}{The recording is of high technical quality, with no distortion or background interference, and the speaker’s voice is captured close to the microphone, ensuring clarity and intelligibility.} At the moment the speaker is about to elaborate further, the audio is abruptly cut off, truncating his final word and leaving the thought incomplete. \textcolor{Red}{This sudden interruption is accompanied by a sharp, digital click, characteristic of a hard edit or technical error, and is immediately followed by a brief, high-frequency electronic tone. The tone is static and unchanging, likely the result of a digital artifact from the editing process, and is not part of the original speech. No other sounds—such as music, ambient noise, or environmental cues—are present, and the absence of audience reactions or other voices indicates a solitary recording context.} In summary, the clip features a single, articulate male speaker with a Caribbean accent discussing a persistent, long-standing idea in an unembellished, contemplative manner. \textcolor{Red}{The audio is clear and focused, set in a quiet, private environment, and is abruptly interrupted by a technical edit, which introduces a digital click and an electronic tone, cutting off the speaker mid-sentence and leaving the discussion unresolved.} \\
\midrule
Audio + User Prompt w/o Tags & A male speaker with a medium-low pitched, slightly gravelly voice and a \textcolor{Red}{non-native English accent} delivers a slow, deliberate, and thoughtful statement. His speech is marked by a slight repetition at the beginning and pauses between phrases, with a steady, calm, and measured pace. The intonation is even and declarative, conveying a sense of reflection and certainty, as if presenting a considered opinion or fact.\\
\midrule
Audio + User Prompt w/ Tags & A mature male with a clear, medium-low pitched voice and a \textcolor{Green}{Jamaican accent} speaks in a measured, deliberate manner. He begins with a slightly rushed, emphatic tone and a moderate pitch, conveying a sense of conviction. After a short pause, he continues at a steady, thoughtful pace with a slightly lower pitch and a more subdued, reflective tone. The speech is delivered at a moderate volume with a controlled, conversational delivery.\\
\bottomrule[1pt]
\end{tabular}
\end{table*}
\begin{table*}[!t]
\vspace{1em}
\small
\centering
\caption{Case study of multi-positive caption generation for the same speech clip using different random seeds.}
\label{tab:case_study_multi_postive_caption}
\renewcommand\tabcolsep{4pt}
\begin{tabular}{p{0.10\linewidth} p{0.86\linewidth}}
\toprule[1pt]
\textbf{Seed} & \textbf{Fine-grained Caption} \\
\midrule
1 & A middle-aged male with a clear, medium-pitched voice and an Australian accent speaks in a conversational, explanatory tone. His speech is delivered at a fast, steady pace, with a moderate volume and smooth, fluid articulation. The intonation is relatively flat and consistent, conveying a matter-of-fact, informative style. \\
\midrule
2 & A male speaker with a medium-low pitched, resonant voice and an Australian accent speaks in a measured, deliberate manner. His delivery is calm and conversational, with a steady rhythm and even pacing, using a moderate volume and a flat, unemotional tone. \\
\midrule
3 & A mature male with a clear, medium-pitched voice and an Australian accent speaks in a conversational, informal style. He begins at a fast, energetic pace with an emphatic, high pitch and moderately loud volume, using a casual, slightly breathy tone. After a brief pause, he continues at a steady, conversational pace with a moderate pitch and volume, his tone remaining expressive and fluctuating throughout the fluent delivery. \\
\midrule
4 & A mature male with a medium-low pitched, resonant voice and a distinct Australian accent speaks in a measured, confident manner. His delivery is steady and deliberate, with a moderate speaking rate and clear articulation. The tone is calm and conversational, suggesting an explanatory or persuasive intent. He maintains a consistent volume and pitch, with subtle inflections that convey a sense of authority and familiarity. \\
\midrule
5 & A mature male with a clear, medium-pitched voice and an Australian accent speaks in a conversational, explanatory style, similar to a commentator or analyst. He begins with a moderate pace and clear enunciation, using a mid-to-low pitch that rises slightly for emphasis. After a short pause, he continues at a slightly faster pace with a more declarative tone, and his pitch drops again as he concludes with a downward inflection, maintaining a steady, conversational delivery throughout. \\
\bottomrule[1pt]
\end{tabular}
\vspace{1em}
\end{table*}

\section{Case Study: User Prompt Composition for Fine-Grained Caption Generation}
\label{sec:user_prompt_detailed_caption}

This case study qualitatively examines how different user prompt compositions influence the fine-grained captions generated by Qwen3-Omni-30B-A3B-Captioner.

\begin{figure*}[th]
\centering
\includegraphics[width=0.95\linewidth]{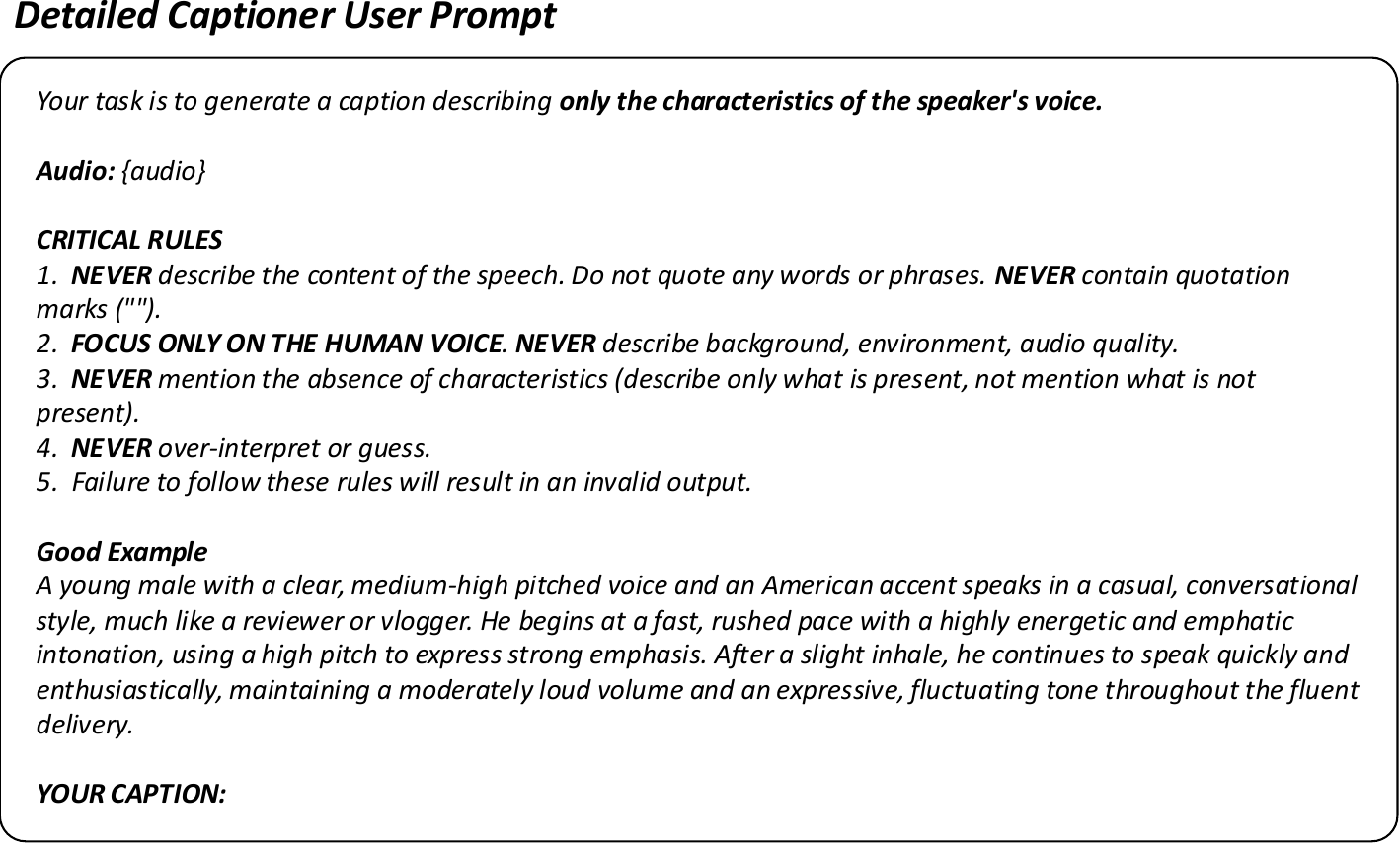}
\caption{User prompt for detailed captioner.}
\label{fig:captioner_prompt}
\end{figure*}

\begin{figure*}[th]
\centering
\includegraphics[width=0.95\linewidth]{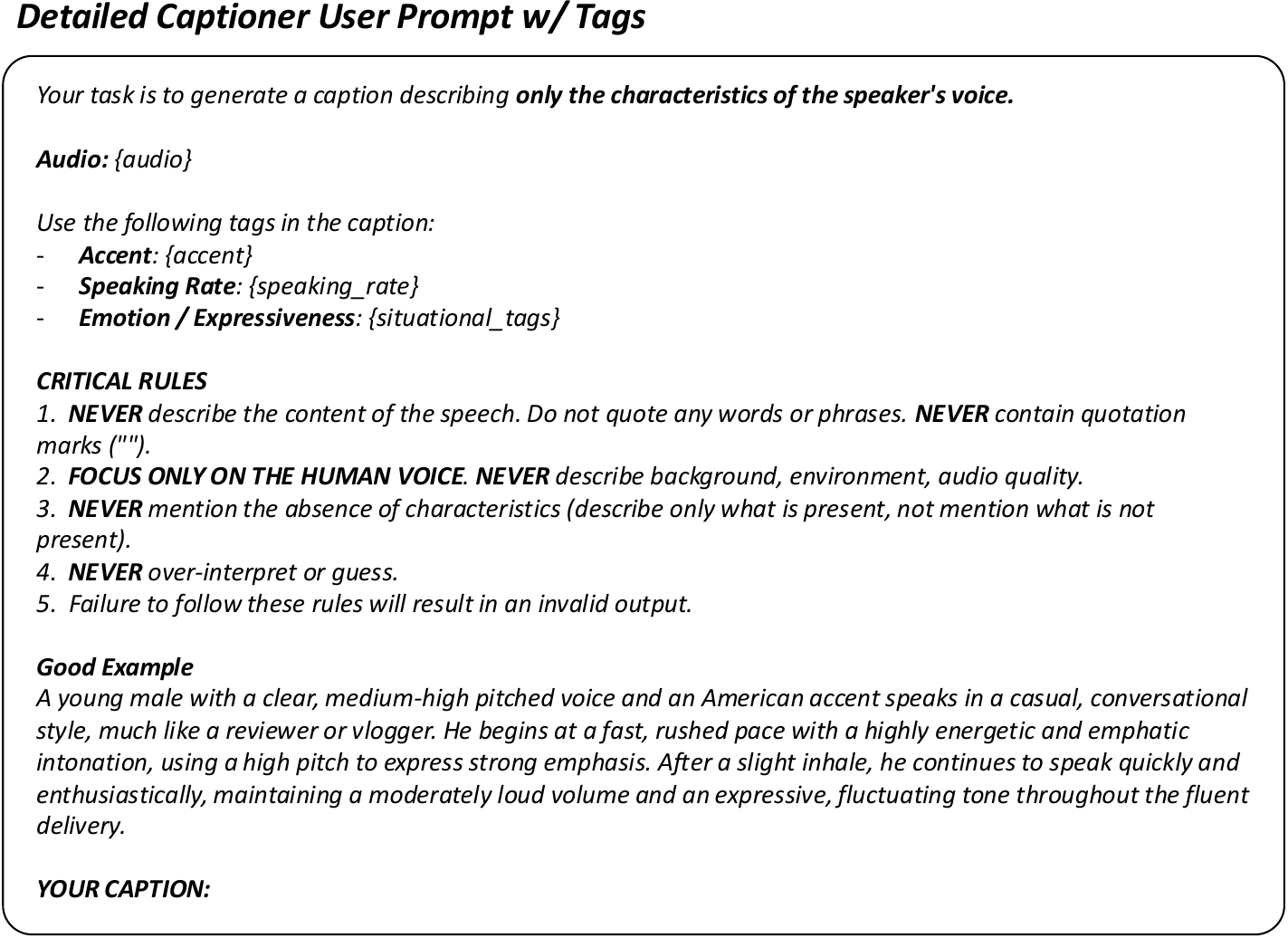}
\caption{User prompt with human-annotated tags for detailed captioner.}
\label{fig:captioner_prompt_w_tags}
\end{figure*}

\paragraph{Setup}
We compare three prompt compositions:
\begin{itemize}[leftmargin=*, itemsep=0pt, topsep=2pt]
\item \textbf{Audio-only}: The captioner takes audio as input and generates captions in its default manner.
\item \textbf{Audio + User Prompt (w/o Tags)}: A user prompt (Figure~\ref{fig:captioner_prompt}) is used to discourage spoken-content transcription, environment-related descriptions, and audio quality assessments.
\item \textbf{Audio + User Prompt (w/ Tags)}: A user prompt with human-annotated attributes (Figure~\ref{fig:captioner_prompt_w_tags}) is used to guide caption generation toward accurate realization of specified attributes, especially those that are rare or inherently ambiguous when inferred from audio alone.
\end{itemize}

\paragraph{Analysis}
As shown in Table~\ref{tab:case_study_user_prompt}, different prompt compositions lead to distinct captioning behaviors.
\begin{itemize}[leftmargin=*, itemsep=0pt, topsep=2pt]
\item When no user prompt is applied, the generated caption includes spoken-content transcription, detailed descriptions of environmental background noise, audio quality assessments, speculative mentions of editing artifacts and electronic tones, as well as explicit declarations about the absence of certain elements, all highlighted in red.
Such speaker-independent and verbose content can distract contrastive learning from speaker-centric representations, encouraging reliance on shallow lexical cues from verbatim transcription or recording-specific artifacts.
\item Introducing a lightweight user prompt results in a more concise and speaker-focused caption.
However, the inferred accent remains coarse, highlighted in red, reflecting the intrinsic ambiguity of accent identification from short speech segments, where many regional accents exhibit highly similar acoustic patterns and are difficult to reliably distinguish based on audio alone.
\item When human-annotated attributes are additionally provided, the specified accent can be realized accurately in the generated caption, highlighted in green. The resulting description not only reflects the correct accent but also maintains a coherent temporal structure that captures changes in speaking style and delivery.
\end{itemize}

Overall, this comparison illustrates how human annotations can resolve ambiguity in inherently hard-to-delineate speaker traits, while preserving the narrative nature of fine-grained captions.

\section{Case Study: Multi-Positive Caption Generation}
\label{sec:multi_postive_caption}

\paragraph{Analysis}
Table~\ref{tab:case_study_multi_postive_caption} presents multiple fine-grained captions generated for the same speech clip using different random seeds.
All captions consistently capture core speaker attributes, including gender, accent, and broadly similar pitch range and speaking rate, indicating that they remain well grounded in the same underlying speech signal.
At the same time, the captions exhibit notable variation in lexical choice, descriptive emphasis, and narrative structure.
Some emphasize overall delivery style and speaker demeanor, while others describe within-utterance dynamics such as changes in pace, pauses, or shifts in intonation.
Together, these captions constitute multiple semantically non-identical textual views of the same speech.

\begin{figure*}[!t]
\centering
\includegraphics[width=\linewidth]{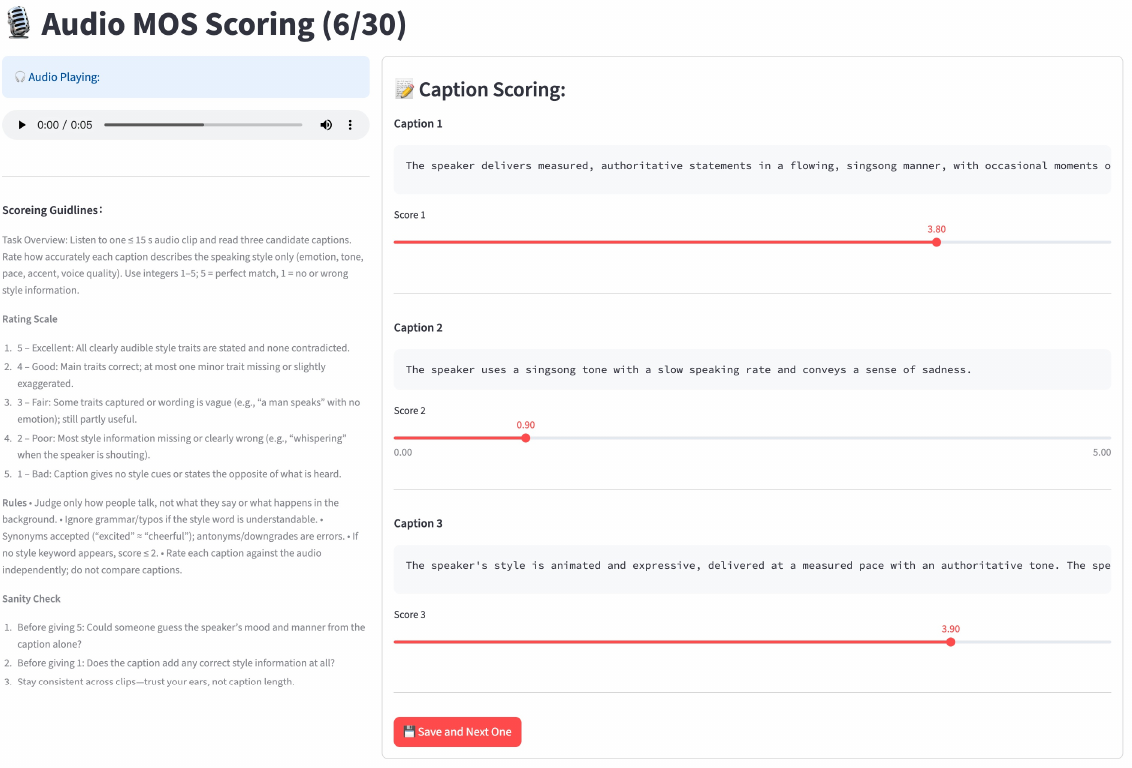}
\vspace{-1em}
\caption{Annotation UI for raters to annotate the alignment score between one audio and several candidate captions.}
\label{fig:mos_page}
\end{figure*}

\section{Detailed Protocol of LLM-as-Judges}
\label{sec:llm_protocol}
\begin{figure*}[th]
\centering
\includegraphics[width=0.95\linewidth]{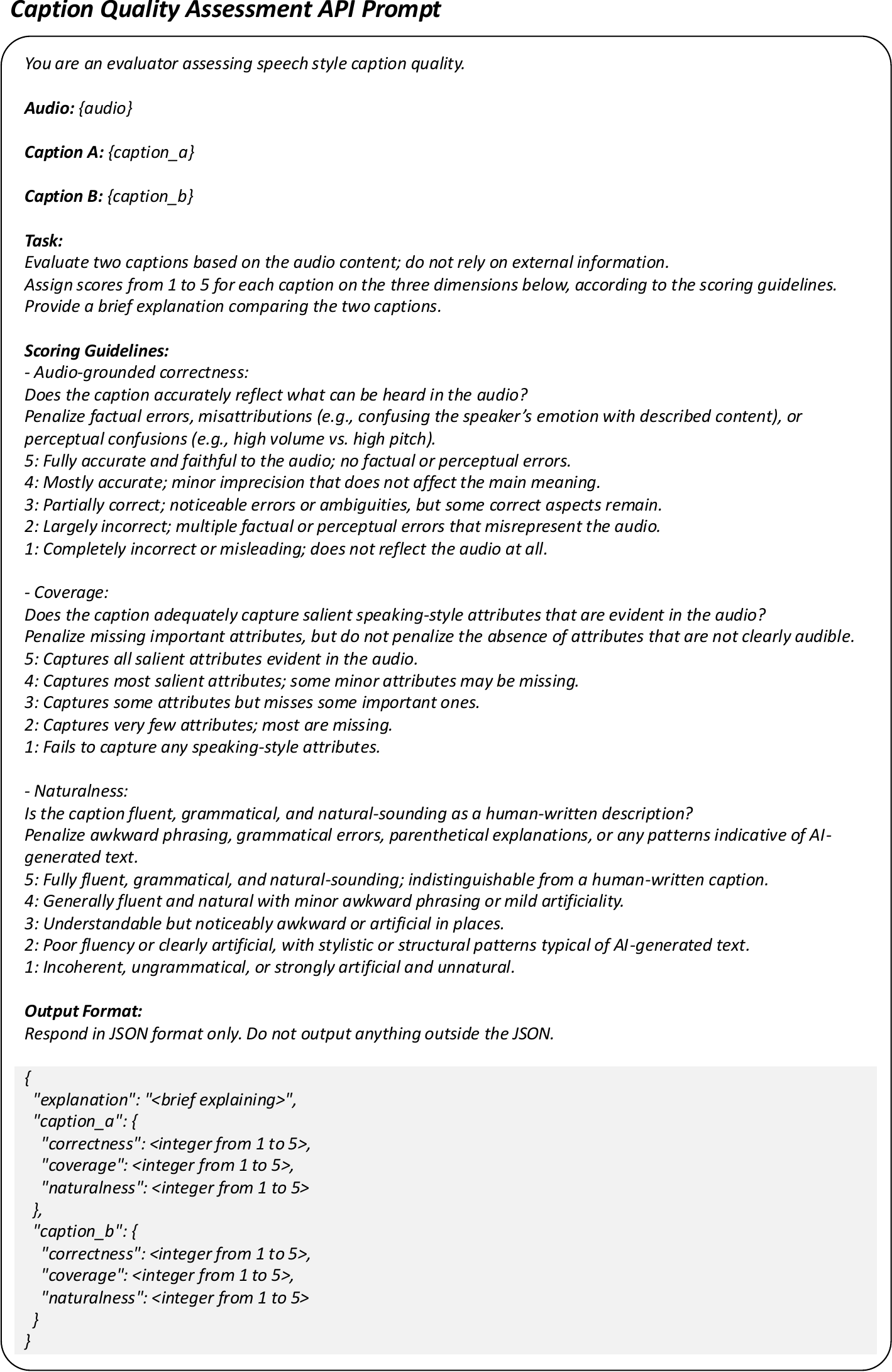}
\caption{Detailed protocol of LLM-as-Judges.}
\label{fig:gemini_prompt}
\end{figure*}

The LLM-as-Judges evaluation follows three criteria, each assessed using a five-level rubric that specifies both failure modes and ideal behaviors, as detailed in Figure~\ref{fig:gemini_prompt}.

\section{Details of Subjective Evaluation}
\label{sec:subjective_eval_detail}

\paragraph{Evaluation Data Selection}
To validate whether \ourmodel{} scores align with human perception, we conduct a meta-evaluation using the evaluation set of ParaSpeechCaps. This dataset contains diverse audio samples paired with style descriptions covering two key paralinguistic dimensions:
\begin{itemize}[leftmargin=*, itemsep=0pt, topsep=2pt]
\item Intrinsic Speaker Traits: Speaker characteristics tied to a speaker’s identity, such as gender, age, timbre/texture, pitch, accent, and so on. 
\item Situational Traits: Dynamic aspects including speaking rate, emotion, expressivity, volume, and other speaker style-related attributes. 
\item Fusion: Complex audio captions that combine both intrinsic and situational descriptors.
\end{itemize}
For each category, we randomly select 30 audio clips for human evaluation and scoring, and adopt a text-based large language model Qwen3-30B-A3B-Instruct-2507\footnote{\url{https://huggingface.co/Qwen/Qwen3-30B-A3B-Instruct-2507}}~\cite{qwen3} to rewrite and manually filter the official captions provided, ensuring that the final captions only contain the intrinsic speaker or situational traits for evaluation.

\paragraph{Human Annotation Protocol}
Figure~\ref{fig:mos_page} illustrates the user interface used for human subjective evaluation.
For each evaluation instance, 20 raters were presented with an audio clip together with three candidate captions describing the intrinsic speaker or situational traits.
Raters were allowed to replay the audio as needed before providing their judgments.
The correlation score was assessed using a Mean Opinion Score (MOS) protocol.
Raters assigned a score to each caption via a slider on a continuous scale from $0.0$ to $5.0$, where higher values indicate greater perceptual similarity and stronger consistency between the caption and the audio.
Detailed scoring guidelines were displayed alongside the interface. Raters were instructed to evaluate each caption independently based on how accurately it captured style attributes of the speech, while ignoring irrelevant factors such as grammatical errors or minor wording variations.

\paragraph{Evaluation Metrics}
We employ three widely recognized statistical coefficients to measure the agreement between our model-derived similarity scores and human subjective ratings:
\begin{itemize}[leftmargin=*, itemsep=0pt, topsep=2pt]
\item Pearson Correlation Coefficient ($r$): Evaluates the linear relationship between the model predictions and human scores.
\item Spearman’s Rank Correlation ($\rho$): Assesses the monotonic relationship, reflecting how well the model preserves the relative ranking of samples.
\item Kendall’s Tau ($\tau$): Measures the ordinal association and pairwise agreement between the two sets of rankings, providing a robust check for consistency.
\end{itemize}

\section{Visualization of Subjective Evaluation}
\label{sec:visual_subjective_evaluation}
\begin{figure*}[!t]
    \centering
    
    \hspace*{0.5em}
    \begin{minipage}{0.32\textwidth} \centering \small Intrinsic Traits \end{minipage} \hfill
    \begin{minipage}{0.32\textwidth} \centering \small Situational Traits \end{minipage} \hfill
    \begin{minipage}{0.32\textwidth} \centering \small Fusion \end{minipage}
    
    \vspace{1em}

    \rotatebox{90}{\quad \quad \quad \small LAION-AI CLAP}
    \includegraphics[width=0.32\textwidth]{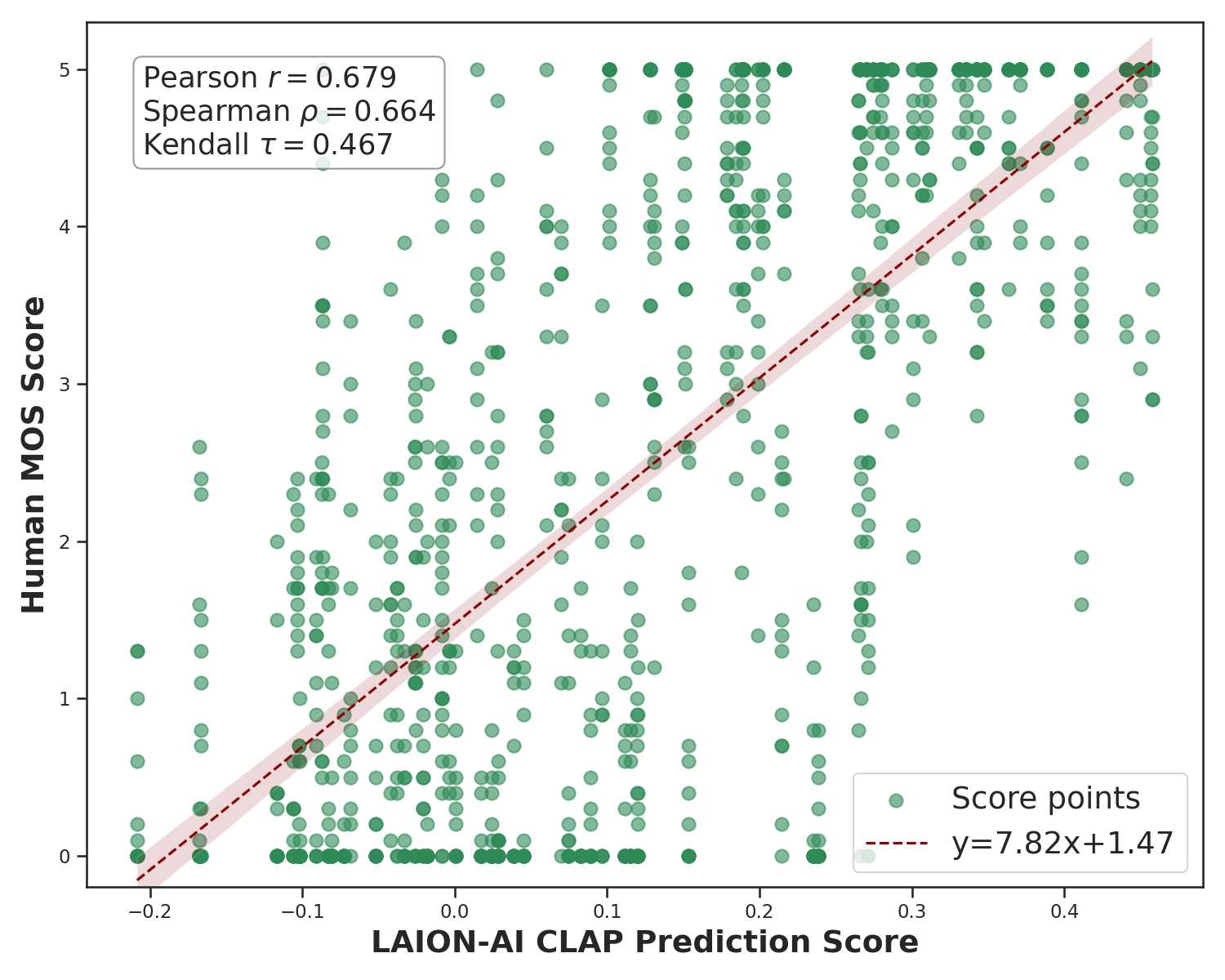} \hfill
    \includegraphics[width=0.32\textwidth]{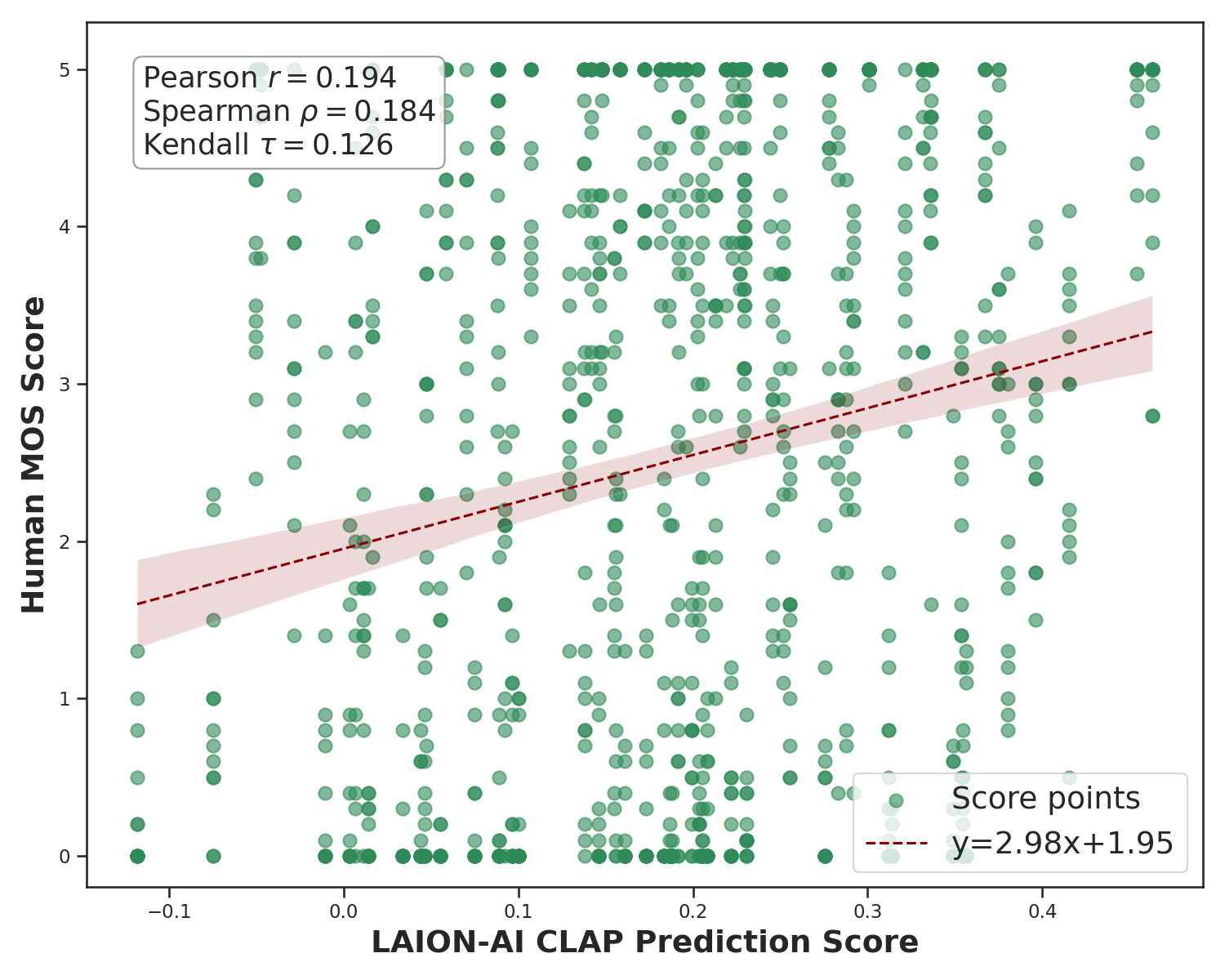} \hfill
    \includegraphics[width=0.32\textwidth]{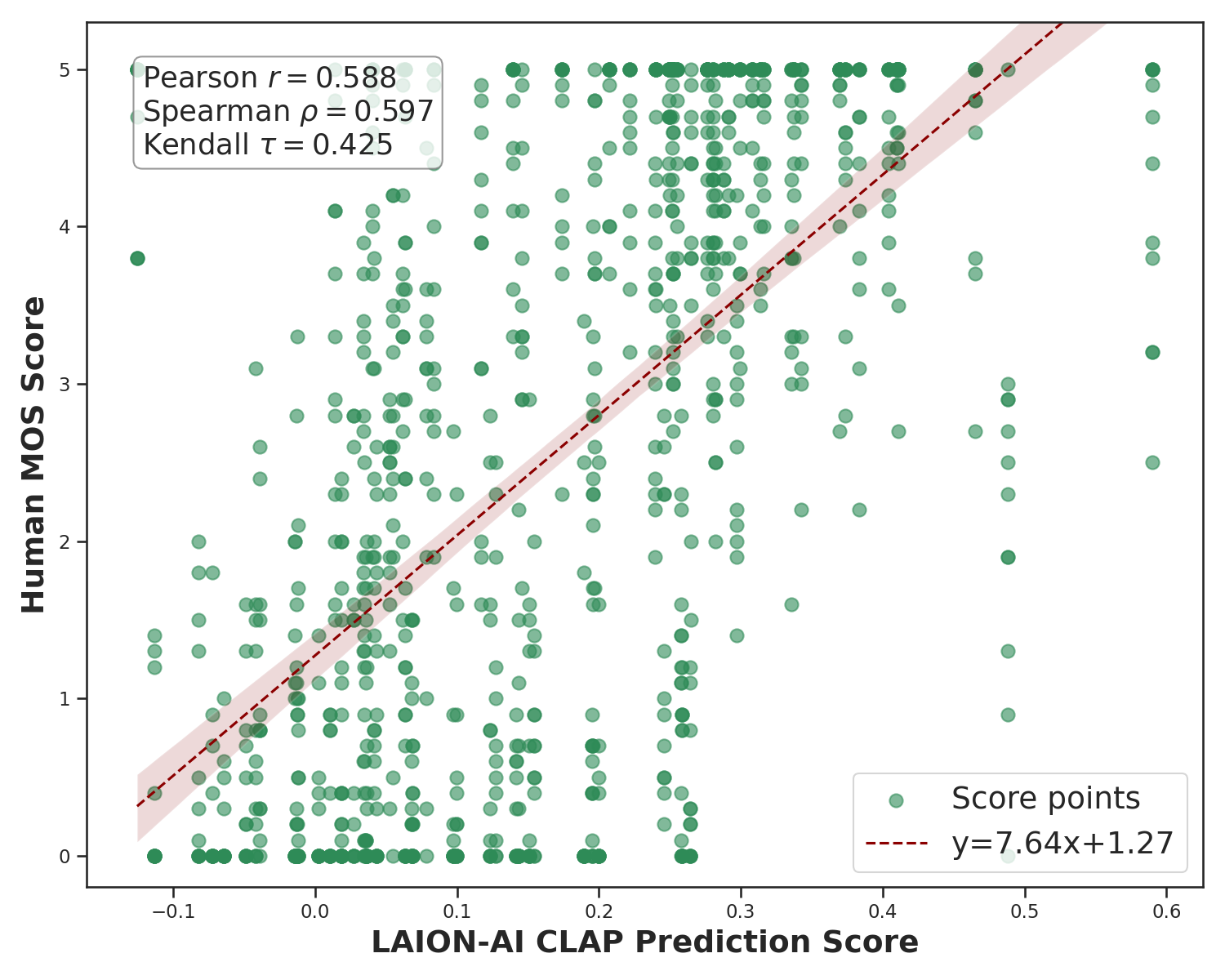} 

    \vspace{1em}

    \rotatebox{90}{\quad \quad \quad \quad \quad \small GLAP}
    \includegraphics[width=0.32\textwidth]{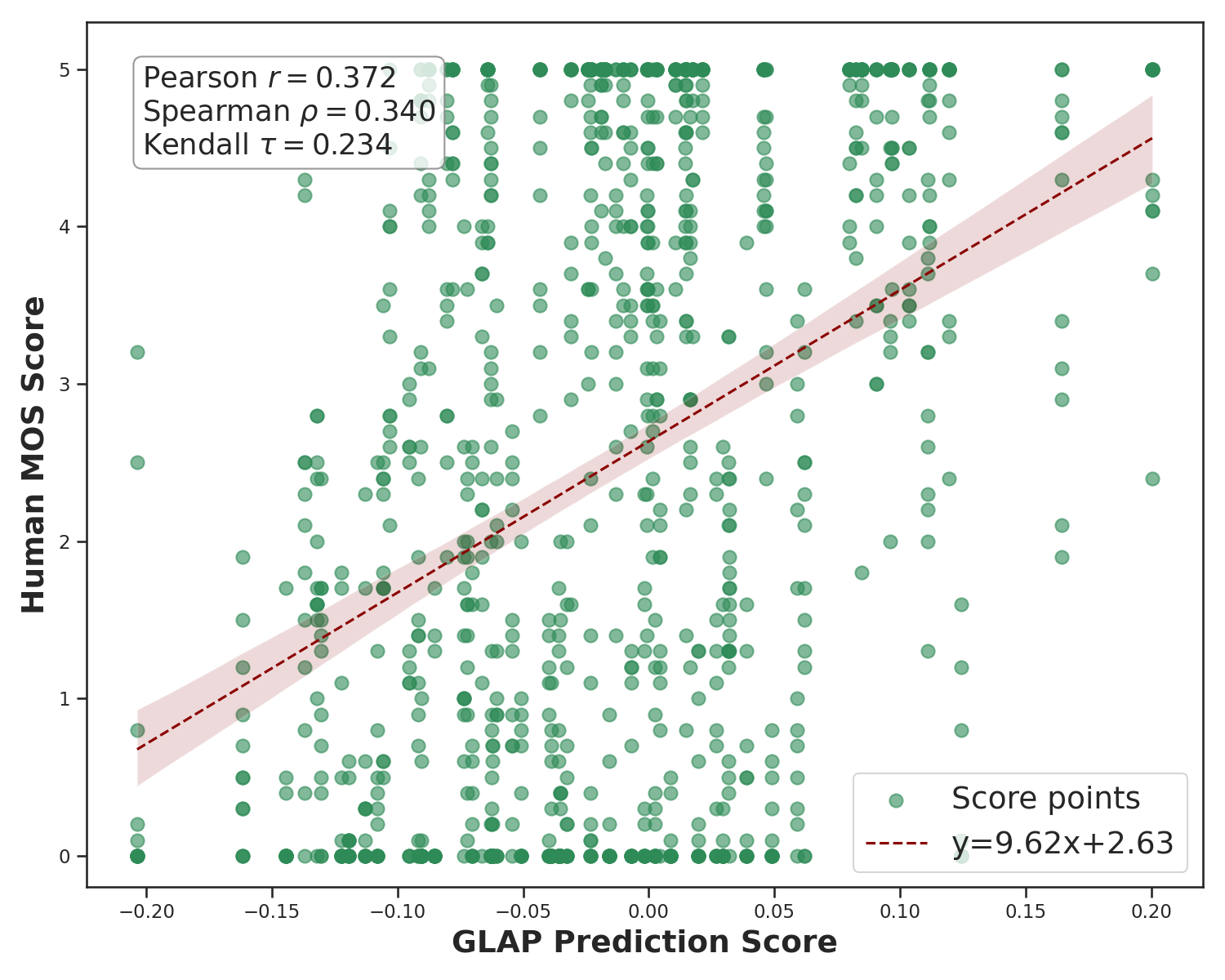} \hfill
    \includegraphics[width=0.32\textwidth]{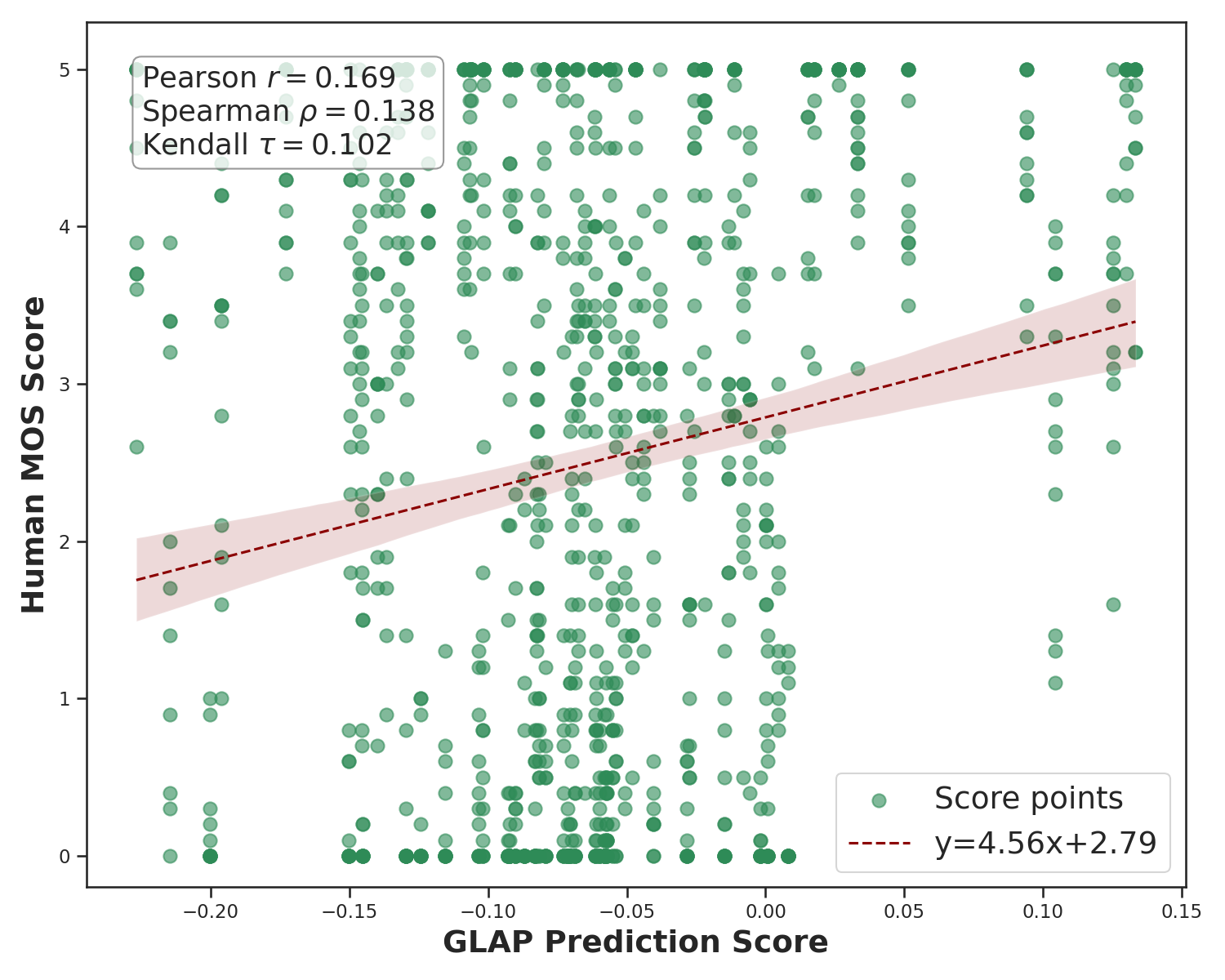} \hfill
    \includegraphics[width=0.32\textwidth]{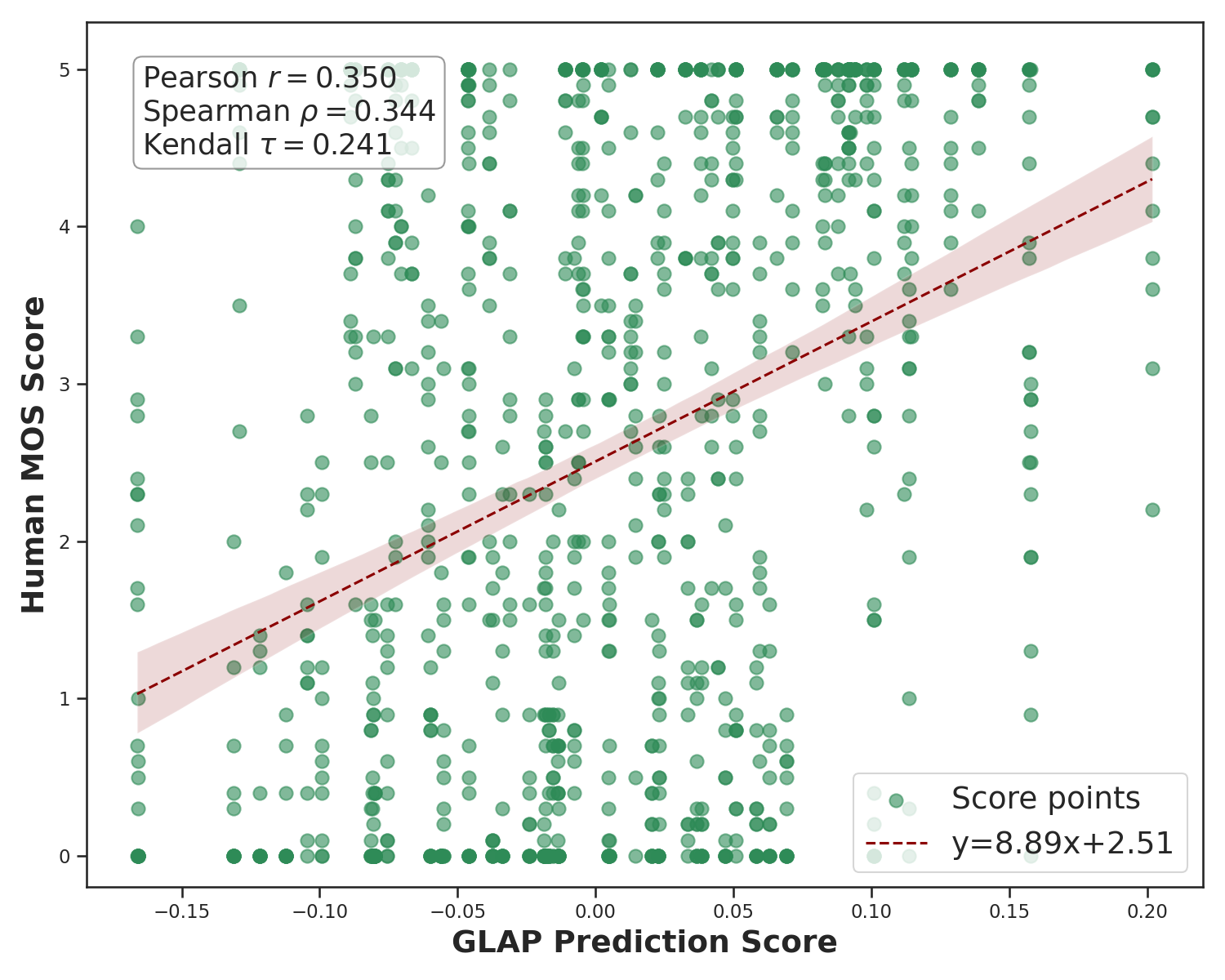} 

    \vspace{1em}

    \rotatebox{90}{\quad \quad \quad \quad \small ParaCLAP}
    \includegraphics[width=0.32\textwidth]{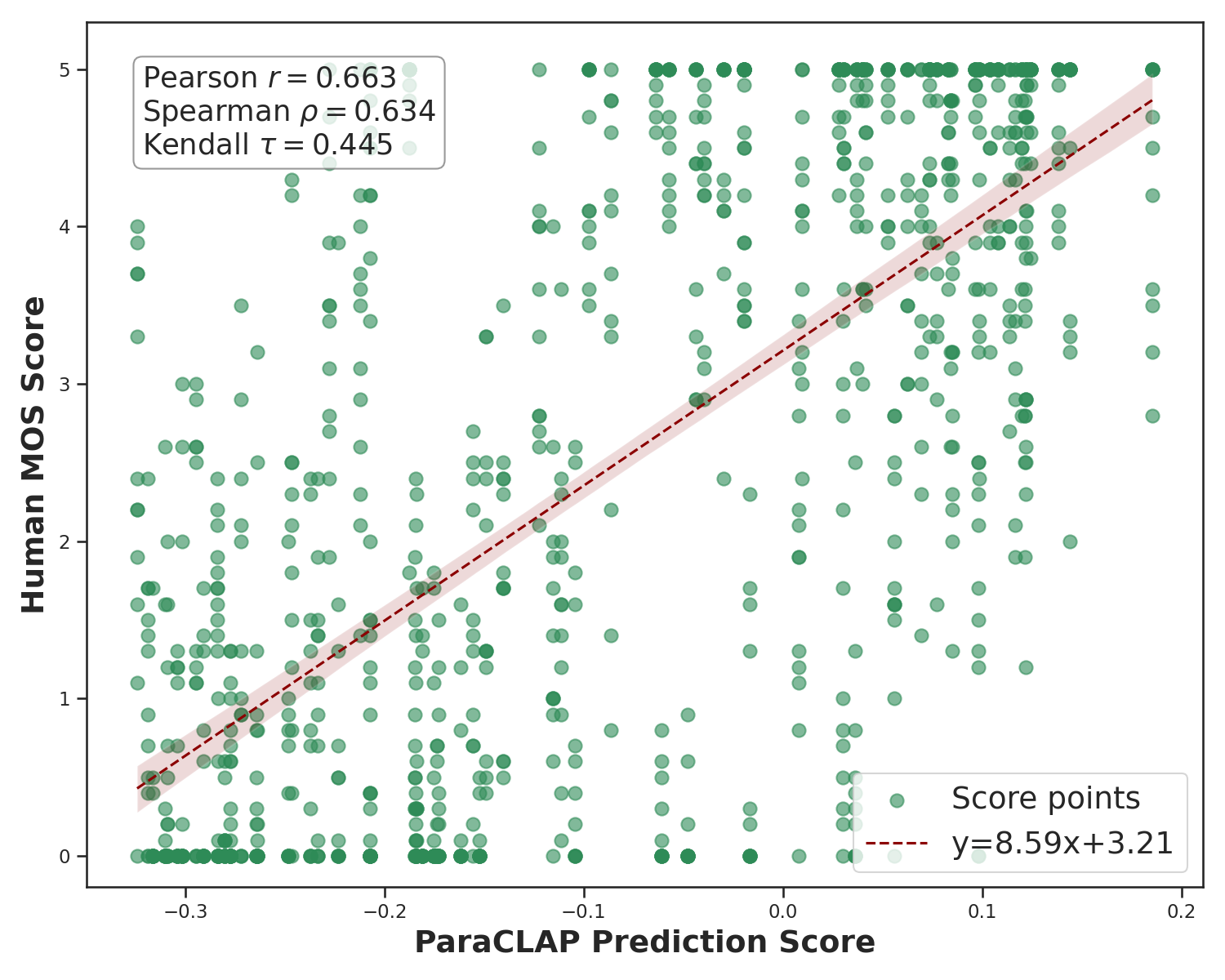} \hfill
    \includegraphics[width=0.32\textwidth]{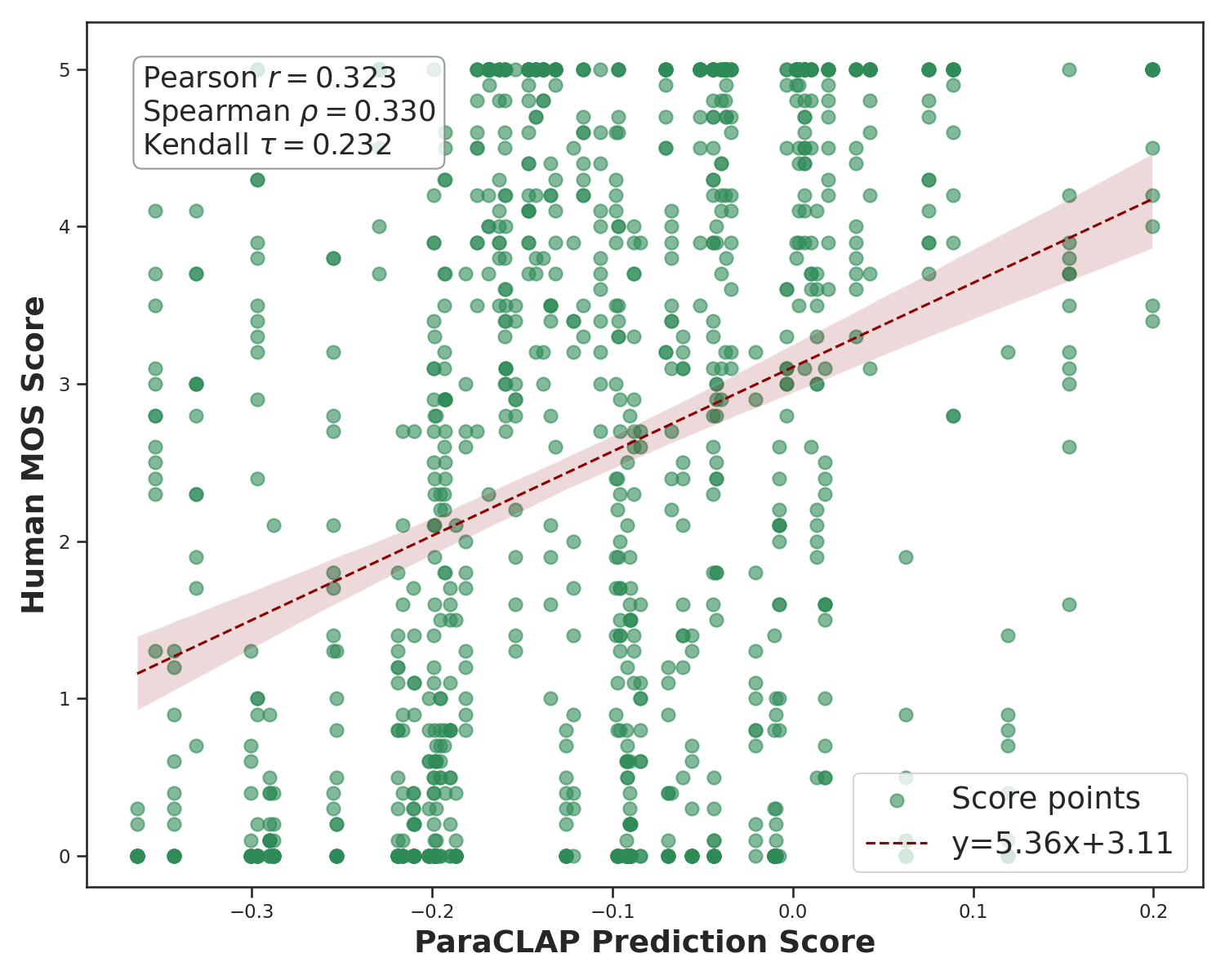} \hfill
    \includegraphics[width=0.32\textwidth]{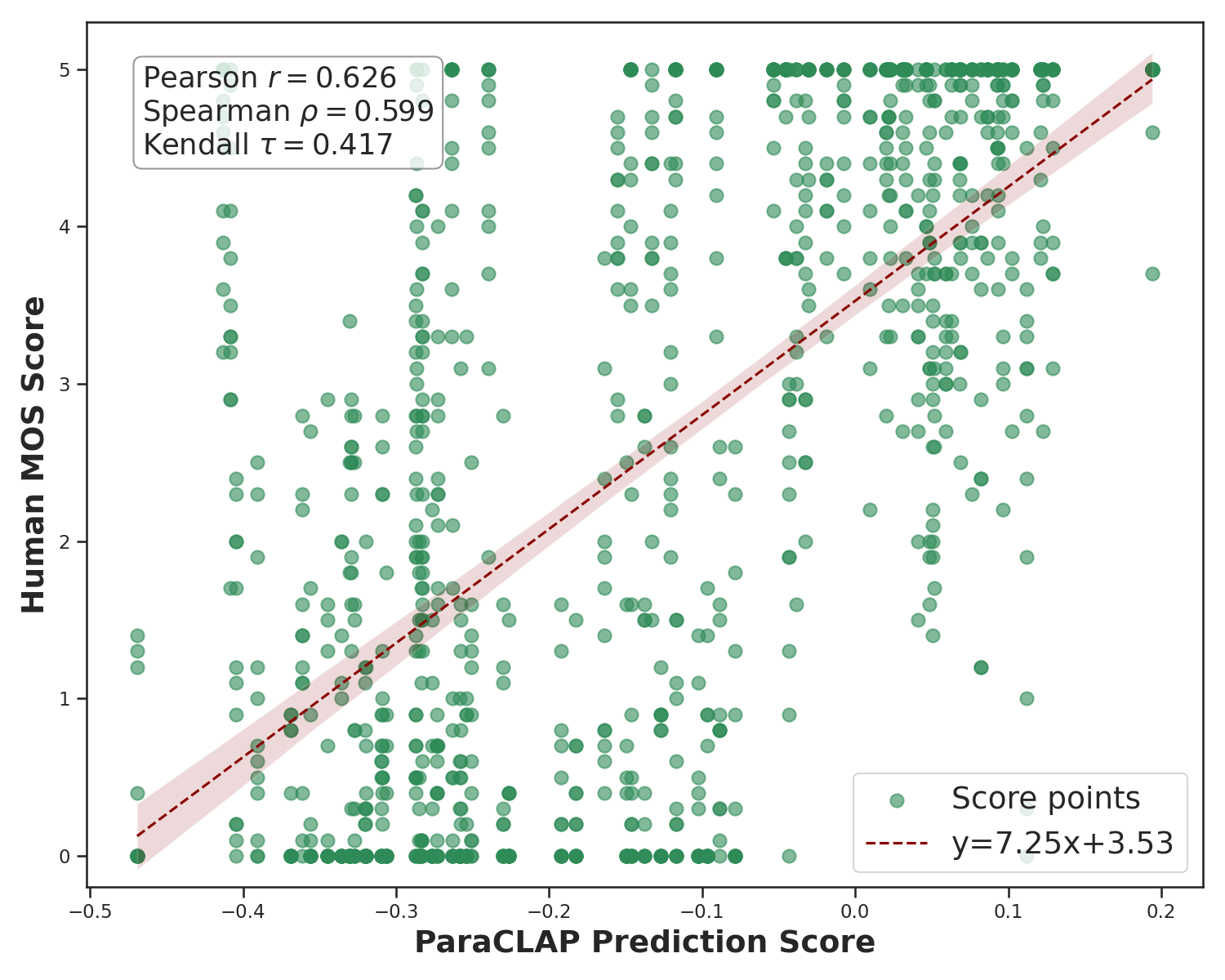} 

    \vspace{1em}

    \rotatebox{90}{\quad \quad \quad \quad \small \ourmodel{}}
    \includegraphics[width=0.32\textwidth]{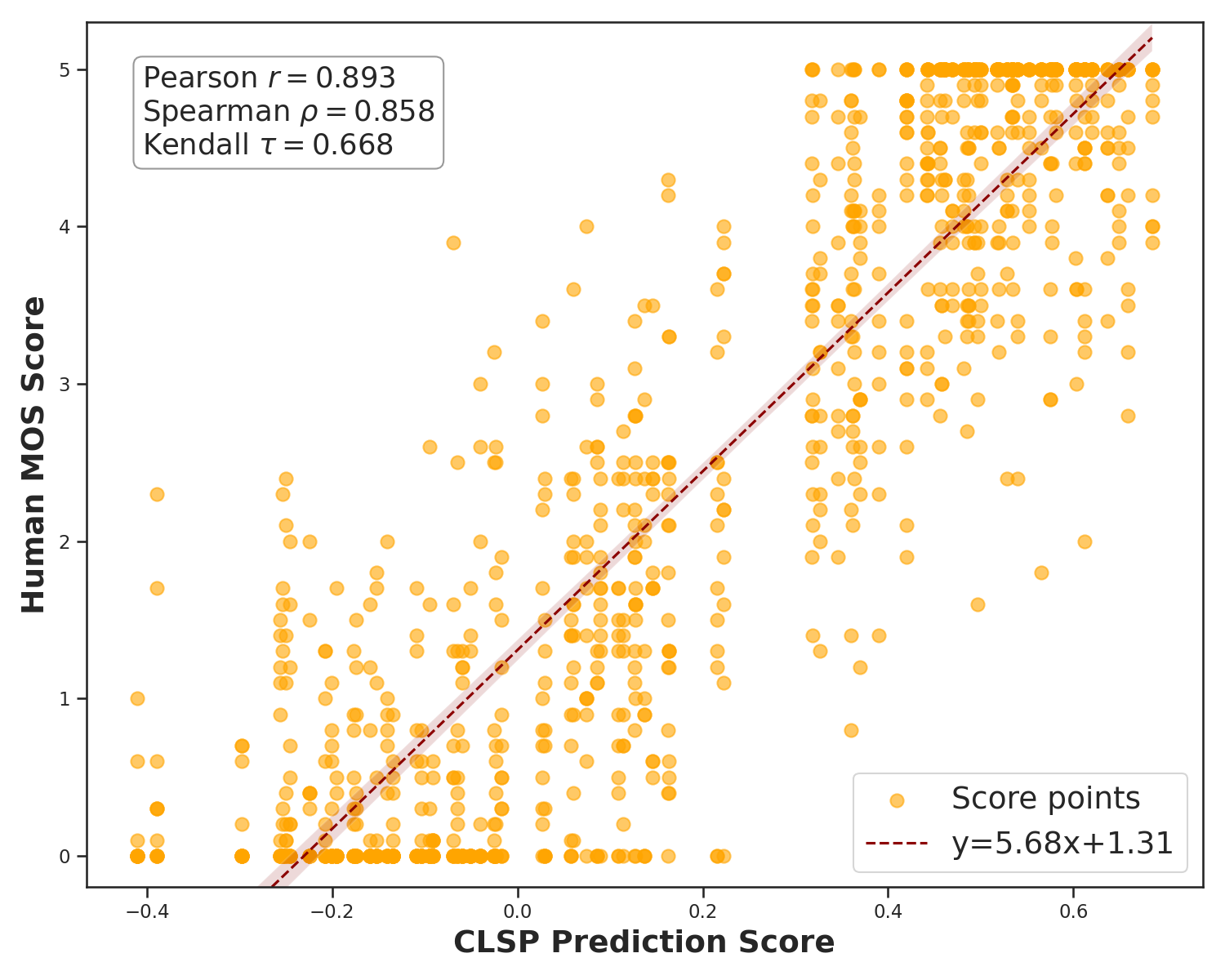} \hfill
    \includegraphics[width=0.32\textwidth]{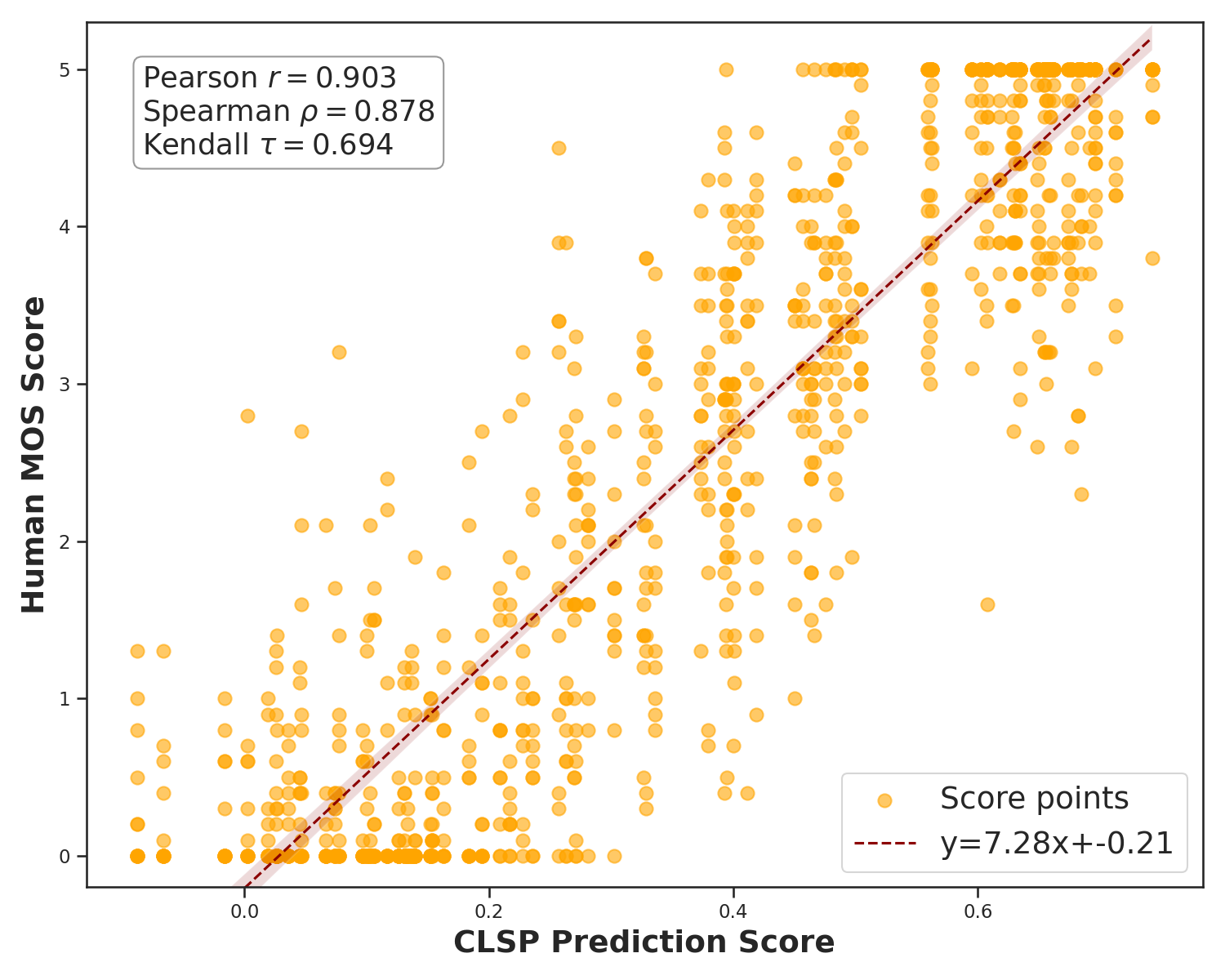} \hfill
    \includegraphics[width=0.32\textwidth]{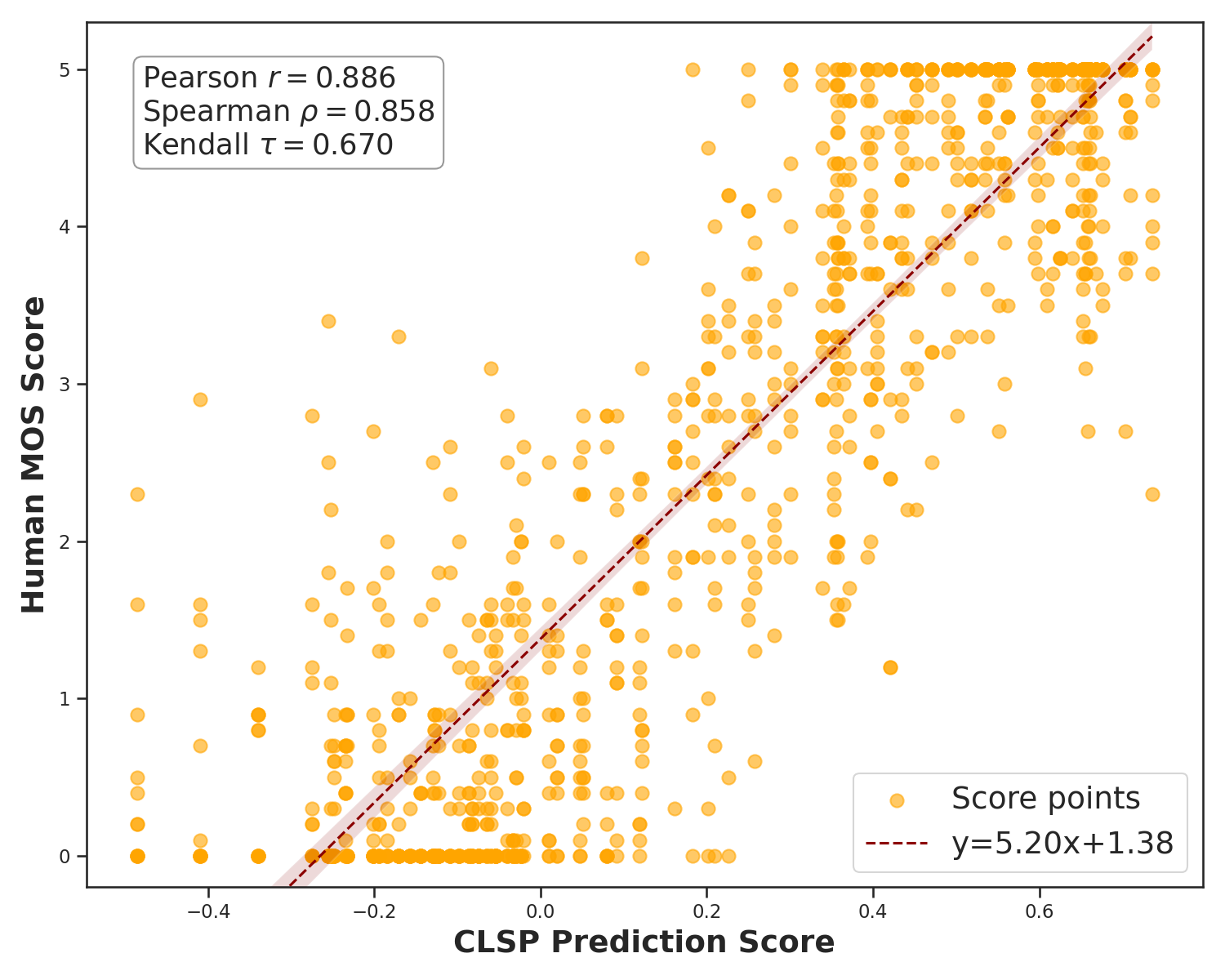} 

    \vspace{1em}

    \caption{Correlation analysis between model-predicted similarity scores and subjective human ratings across different models and trait categories (Intrinsic, Situational, and Fusion). The green scatter plots demonstrate the alignment between automated metrics and human perception using the ParaSpeechCaps evaluation set. The dashed red lines represent the linear regression fit.}
    \label{fig:visual_scatter}
\end{figure*}

Figure~\ref{fig:visual_scatter} provides a visual comparison of the correlation between model-predicted similarity scores and human subjective ratings across four models and three paralinguistic trait categories. Compared to baseline models, the scatter plots for \ourmodel{} (bottom row) exhibit a significantly tighter clustering of data points around the red linear regression lines. This pattern is consistent across all sub-categories. The higher density of points along the diagonal confirms that \ourmodel{} can reliably replicate the nuanced judgments of human experts. 

The visual trends observed in Figure~\ref{fig:visual_scatter}, together with the strong Pearson, Spearman, and Kendall’s Tau correlations reported in Table~\ref{tab:sub_score}, suggest that \ourmodel{} provides a reliable automated proxy for assessing speech-text style alignment.
In particular, \ourmodel{} captures relative ranking and preference patterns that are consistent with human judgments, supporting its use for large-scale evaluation in speech-text matching scenarios where human annotation is costly.

\section{Ablation Studies}
\label{sec:ablation}
We conduct ablation studies on the following components to provide an in-depth understanding of the training of \ourmodel{}:
\begin{itemize}[leftmargin=*, itemsep=0pt, topsep=2pt]
    \item \textbf{Multi-stage training}: We compare models trained with different stages (see Appendix~\ref{sec:ablation_multi_stage_training}).
    \item \textbf{Target weight}: We compare different target weights $\lambda$ in multi-positive InfoNCE loss used in Stage Two (see Appendix~\ref{sec:ablation_target_weight}).
    \item \textbf{Task scheduler}: We compare different task scheduling strategies used in Stage Two (see Appendix~\ref{sec:ablation_task_scheduler}).
\end{itemize}

\subsection{Ablation Study: Multi-Stage Training}
\label{sec:ablation_multi_stage_training}
Table~\ref{tab:ablation_training_stage} presents an ablation study examining the effectiveness of each training stage.
Training with only Stage~1 primarily benefits fine-grained speech--text retrieval, whereas training with only Stage~2 achieves only limited performance on both global and fine-grained retrieval tasks.
Combining both stages yields the best performance across all metrics, demonstrating that each stage contributes effectively to both global and fine-grained speech--text alignment.

\subsection{Ablation Study: Target Weight}
\label{sec:ablation_target_weight}
Table~\ref{tab:ablation_target_weight} reports an ablation study on the loss weight $\lambda$.
Performance remains stable across a broad range of $\lambda$ values, indicating that the proposed training objective is not overly sensitive to the exact choice of the target weight.
Among the tested settings, $\lambda=0.5$ achieves the best overall performance on both global and fine-grained speech--text alignment, and is therefore used in all experiments.

\subsection{Ablation Study: Task Scheduler}
\label{sec:ablation_task_scheduler}
Table~\ref{tab:ablation_task_scheduler} presents an ablation study of different task scheduling strategies.

For static mixtures, decreasing the sampling probability of Task~1 shifts the training focus towards fine-grained discrimination via semantic consistency, leading to consistent improvements in fine-grained retrieval performance but a notable degradation in global retrieval, particularly when $p_0$ falls below 0.2.
Conversely, assigning a higher sampling probability to Task~1 improves both global and fine-grained performance by encouraging effective cross-granularity generalization, although fine-grained retrieval does not reach the level achieved when training exclusively with Task~2.
These reveal a trade-off in static scheduling strategies.

In contrast, dynamic schedulers effectively combine the advantages of both tasks and achieve overall stronger and more balanced performance across global and fine-grained retrieval.
The best results are obtained by a dynamic scheduler that gradually shifts the sampling distribution from Task~1 to Task~2, with $p_0=0.95$, $p_{\min}=0.50$, and $T=10{,}000$, confirming the effectiveness of curriculum-style task scheduling.

\begin{table*}[th]
\small
\centering
\caption{Ablation study of different training stages, evaluated on global speech--text retrieval and fine-grained speech--text retrieval tasks. \ding{51} / \ding{55} indicate whether a training stage is performed.}
\label{tab:ablation_training_stage}
\renewcommand\tabcolsep{8pt}
\begin{tabular}{ccccccc}
\toprule[1pt]
\multirow{3}{*}{\textbf{Stage 1}}
& \multirow{3}{*}{\textbf{Stage 2}} 
& \multicolumn{2}{c}{\textbf{Global Speech--Text Retrieval}} 
& \multicolumn{2}{c}{\textbf{Fine-Grained Speech--Text Retrieval}}
& \multirow{3}{*}{\textbf{Avg.}} \\
\cmidrule(lr){3-4} \cmidrule(lr){5-6}
& & Speech-to-Text & Text-to-Speech & Speech-to-Text & Text-to-Speech \\
& & mAP@10 & mAP@10 & mAP@10 & mAP@10 \\
\midrule
\ding{51} & \ding{55} & 11.1 & 9.6 & 69.6 & 64.0 & 38.6 \\
\ding{55} & \ding{51} & 38.3 & 38.5 & 35.3 & 32.6 & 36.2 \\
\ding{51} & \ding{51} & \textbf{58.7} & \textbf{54.5} & \textbf{77.9} & \textbf{77.2} & \textbf{67.1}\\
\bottomrule[1pt]
\end{tabular}
\end{table*}
\begin{table*}[th]
\small
\centering
\caption{Ablation study of different loss weights, evaluated on global speech--text retrieval and fine-grained speech--text retrieval tasks.}
\label{tab:ablation_target_weight}
\renewcommand\tabcolsep{8pt}
\begin{tabular}{cccccc}
\toprule[1pt]
\multirow{3}{*}{\textbf{$\lambda$}}
& \multicolumn{2}{c}{\textbf{Global Speech--Text Retrieval}} 
& \multicolumn{2}{c}{\textbf{Fine-Grained Speech--Text Retrieval}}
& \multirow{3}{*}{\textbf{Avg.}} \\
\cmidrule(lr){2-3} \cmidrule(lr){4-5}
& Speech-to-Text & Text-to-Speech & Speech-to-Text & Text-to-Speech \\
& mAP@10 & mAP@10 & mAP@10 & mAP@10 \\
\midrule
0.3 & 53.4 & 52.6 & \textbf{77.9} & 77.0 & 65.2\\
0.4 & 56.0 & \textbf{55.1} & 77.5 & 75.5 & 66.0\\
0.5 & \textbf{58.7} & 54.5 & \textbf{77.9} & \textbf{77.2} & \textbf{67.1}\\
0.6 & 56.4 & 55.0 & 77.7 & 76.9 & 66.5\\
0.7 & 57.8 & 54.4 & 77.3 & 75.6 & 66.3\\
\bottomrule[1pt]
\end{tabular}
\end{table*}

\begin{table*}[th]
\small
\centering
\caption{Ablation study of different task schedulers, evaluated on global speech--text retrieval and fine-grained speech--text retrieval tasks.}
\label{tab:ablation_task_scheduler}
\setlength{\tabcolsep}{8pt}
\begin{tabular}{rrr cc cc c}
\toprule[1pt]
\multirow{3}{*}{\textbf{$p_0$}} &
\multirow{3}{*}{\textbf{$p_{\min}$}} &
\multirow{3}{*}{\textbf{$T$}} &
\multicolumn{2}{c}{\textbf{Global Speech--Text Retrieval}} 
& \multicolumn{2}{c}{\textbf{Fine-Grained Speech--Text Retrieval}}
& \multirow{3}{*}{\textbf{Avg.}} \\
\cmidrule(lr){4-5} \cmidrule(lr){6-7}
&&& Speech-to-Text & Text-to-Speech & Speech-to-Text & Text-to-Speech \\
&&& mAP@10 & mAP@10 & mAP@10 & mAP@10 \\
\midrule
\multicolumn{8}{l}{\hspace{-2mm}\emph{Static Mixture}} \\
0.00 & -- & -- & 20.4 & 19.0 & 84.9 & 79.6 & 51.0 \\
0.20 & -- & -- & 42.7 & 41.1 & 83.9 & 78.9 & 61.7 \\
0.30 & -- & -- & 51.6 & 49.2 & 81.3 & 76.4 & 64.6 \\
0.40 & -- & -- & 51.1 & 50.6 & 79.2 & 77.1 & 64.5 \\
0.50 & -- & -- & 52.1 & 51.8 & 79.6 & 77.4 & 65.2 \\
0.60 & -- & -- & 54.4 & 52.5 & 79.7 & 76.7 & 65.8 \\
0.70 & -- & -- & 55.8 & 52.9 & 78.8 & 77.9 & 66.4 \\
0.80 & -- & -- & 56.9 & 54.5 & 77.7 & 76.0 & 66.3 \\
0.90 & -- & -- & 57.6 & 56.5 & 77.5 & 75.7 & 66.8 \\
1.00 & -- & -- & 57.6 & 55.9 & 77.9 & 74.9 & 66.6 \\
\midrule
\multicolumn{8}{l}{\hspace{-2mm}\emph{Dynamic Mixture}} \\
0.95 & 0.05 & 5000 & 53.5 & 53.1 & 79.9 & 75.5 & 65.5 \\
0.95 & 0.50 & 5000 & 55.7 & 54.4 & 78.0 & 76.9 & 66.3 \\
0.95 & 0.50 & 10000 & \textbf{58.7} & 54.5 & 77.9 & 77.2 & \textbf{67.1}\\
0.95 & 0.50 & 15000 & 55.6 & 55.5 & 78.2 & 76.1 & 66.4\\

\bottomrule[1pt]
\end{tabular}
\end{table*}
\clearpage

\end{document}